# A mechano-chemical feedback underlies co-existence of qualitatively distinct cell polarity patterns within diverse cell populations

JinSeok Park[1,2,3], William R. Holmes[4], Sung-Hoon Lee[1,2,3], Hong-Nam Kim[5], Deok-Ho Kim[6], Moon Kyu Kwak[7], Chiaochun Joanne Wang[3], Kahp-Yang Suh[8], Leah Edelstein-Keshet[9], Andre Levchenko[1,2,3*]

[1]Department of Biomedical Engineering, Yale University, New Haven, CT 06520, USA.

[2]Yale Systems Biology Institute, Yale University, West Haven, CT 06516, USA

[3]Departments of Biomedical Engineering, The Johns Hopkins University School of Medicine, Baltimore, MD 21205, USA.

[4]Department of Physics and Astronomy, Vanderbilt University, Nashville, TN 37212

[5]Center for BioMicrosystems, Brain Science Institute, Korea Institute of Science and Technology, Seoul, Republic of Korea

[6]Department of Bioengineering, University of Washington, Seattle, WA 98195, USA.

[7]School of Mechanical Engineering, Kyungpook National University, Daegu, Republic of Korea

[8]Department of Mechanical & Aerospace Engineering, Seoul National University, Seoul, Republic of Korea

[9]Department of Mathematics, University of British Columbia, Vancouver, Canada

*To whom correspondence should be addressed: Andre Levchenko, Ph.D. (andre.levchenko@yale.edu).

## ABSTRACT

Cell polarization and directional cell migration can display random, persistent and oscillatory dynamic patterns. However, it is not clear if these polarity patterns can be explained by the same underlying regulatory mechanism. Here, we show that random, persistent and oscillatory migration accompanied by polarization can simultaneously occur in populations of melanoma cells derived from tumors with different degrees of aggressiveness. We demonstrate that all these patterns and the probabilities of their occurrence are quantitatively accounted for by a simple mechanism involving a spatially distributed, mechano-chemical feedback coupling the dynamically changing extracellular matrix (ECM)-cell contacts to the activation of signaling downstream of the Rho-family small GTPases. This mechanism is supported by a predictive mathematical model and extensive experimental validation, and can explain previously reported results for diverse cell types. In melanoma, this mechanism also accounts for the effects of genetic and environmental perturbations, including mutations linked to invasive cell spread. The resulting mechanistic understanding of cell polarity quantitatively captures the relationship between population variability and phenotypic plasticity, with the potential to account for a wide variety of cell migration states in diverse pathological and physiological conditions.

# INTRODUCTION

Cell migration involves complex interactions with the extracellular matrix (ECM) (1-4). Beyond providing cells with the substratum for adhesion and traction during the migration process, the ECM can activate signaling networks through biochemical engagement of the integrin complexes within focal adhesions (FAs) (5-7). The signaling pathways activated by integrins can impinge on the Rho-family small GTPases that are thought to be central regulators of cell polarity and migration (8-10). Varying ECM density can differentially control activation of two proteins belonging to this family, Rac1 and RhoA, which frequently display antagonistic interactions (11, 12). Regulation of Rac1 and RhoA can in turn regulate the mechanical properties of the cell, thus influencing how the cell interfaces with complex local organization of ECM fibers (13-15). The intricate nature of this feedback and the wide diversity of topographic ECM structures have made comprehensive analysis of the resulting cell migration behavior very challenging.

*In vivo* cell migration can display diverse dynamic patterns. It can vary from random exploratory migration characterized by poor FA formation, frequent pseudopod extension and a lack of stress fibers in dense 3D ECM (so-called 3D cell migration) to highly persistent migration along single ECM fibers in sparse 3D ECM (essentially 1D cell migration) (16-18). Recently, oscillatory migration patterns have also been observed following perturbation of cytoskeletal components, with cells retracing their positions over multiple cycles (19-21). Local alignment of ECM fibers can further enhance the persistence and speed of directional cell migration (22). Despite the clear indications that ECM organization is instrumental in eliciting these cell locomotion patterns, we still lack a complete understanding of the underlying molecular mechanisms. In particular, it is not clear if diverse modes of cell migration are all parts of the same spectrum, and whether this spectrum can be affected by intrinsic/extrinsic changes accompanying a transition to more aggressive and invasive cell locomotion *in vivo*.

Genetic alterations can lead to profound changes in both cell-matrix interaction and the propensity for extensive cell migration. One of the most striking examples of such a dramatic alteration in cell behavior is the onset of invasive cell migration within the context of many aggressive cancers (23-25). For instance, the transition to invasive, vertical growth in melanoma is accompanied by a range of mutations, some of which can directly drive invasion (e.g., mutations in integrins) or alter ECM (e.g., mutations affecting matrix metalloproteinases) (26-29). The role of other mutations is less clear; for example, although the loss of phosphatase and tensin homolog (PTEN) is frequently correlated with enhanced melanoma invasiveness (30, 31), the exact nature of how the corresponding change in the signaling network activity might affect invasive cell migration is not known.

In this study, we analyzed how the genetic cell makeup and the organization of ECM could jointly control the patterns of cell migration. We used an experimental model of highly controlled graded changes in the ECM topography in the context of metastatic or non-invasive melanoma cell lines to show that these cells can display a range of well defined migration patterns, related to the spatial organization of intracellular PI3K signaling. In particular, an increase in the anisotropic nature of the ECM organization converted essentially random cell polarity and migration to oscillatory and then to persistent patterns, as cell migration persistence correspondingly increased. Strikingly, a simple mechanism relying on mechano-chemical feedback coupling the dynamically changing extracellular matrix (ECM)-cell contacts to the activation of signaling downstream of Rac1 and RhoA could explain this response and outcomes of additional perturbations of the underlying signaling network. In particular, the model predicted and experiments confirmed that the loss of PTEN, which frequently accompanies transition to more aggressive, invasive cell migration *in vivo*, results in a dramatic increase in spatial stabilization of intracellular signaling and persistence of cell migration. We argue that the mechanisms identified in this analysis elucidate the long observed complex dependencies of cell migration on ECM organization, and can further inform our understanding of invasive cell migration.

**RESULTS**

**Diverse dynamical cell polarity patterns can co-exist, with their relative occurrence modulated by topographic cues**

Cell migration can display diverse patterns, varying in degree of directional persistence (32). The modes of cell locomotion strongly depend on organization and density of the extracellular matrix (ECM) (2, 33, 34), but it is not clear if a gradual change in these ECM features precipitates a commensurately graded and predictable change in the cell migration persistence. To explore this in quantifiable detail, we investigated polarization of cell signaling and patterns of cell migration on surfaces with graded texture. In particular, we varied the density of the nano-scale posts protruding from the cell adhesion surfaces ('cell adhesion substrata'), thus creating a complex extracellular environment presenting cells with nano-topographically complex interface analogous to ECM structure. On such surfaces, if cells are capable of enveloping individual posts, the extent of the overall cell-substratum contact can be much higher than would occur on flat surfaces, modulating the chemical and mechanical input that cells can receive from the ECM coated surfaces (35). The post distribution varied from essentially isotropic to highly anisotropic, gradually increasing along one of the orthogonal axes (*x*-axis). The density along a second orthogonal axis (*y*-axis) was constant (Fig. 1a,b) (For specific dimensions of the surface, see Materials and Methods) (35). We used this system of *x-y* coordinates in further analysis.

We used these cell adhesion substrata to culture an aggressive melanoma cell line, 1205Lu, constitutively expressing GFP or RFP tagged pleckstrin homology domain of Akt (Akt-PH), a common readout of the activity of PI3K. This kinase has been extensively implicated in regulating cell migration, and is usually localized to the front of a motile cell. We found that spatial PI3K signaling, quantified by signaling vector described in Fig. 1c, displayed three distinct dynamic spatial intracellular localization patterns (Fig. 1d-f). The first was characterized by apparently random (RD) localization of PI3K activity 'hot spots' over the course of 3 hours around the whole cell periphery (Fig. 1d and Supplementary movie 1). The second PI3K activity pattern was oscillatory (OS) (Fig. 1e and Supplementary movie 2), with the PI3K signaling persistently alternating between the two sides of the cell along the y-axis of the pattern, with an approximately 30 min. period. Finally, the remaining subset of cells displayed a signaling pattern that was characterized by persistent localization (PS) of the PI3K signaling hot spots to one side of the cell only, along the direction of the *y*-axis (Fig 1f and supplementary movie 3). These results suggested that cell interaction within topographically complex cell-ECM interfaces can lead to diverse PI3K signaling profiles.

Given the graded density of the topography features and concomitant substratum anisotropy, we explored whether and how the three signaling patterns depend on the local substratum topography. To facilitate this analysis, we subdivided the substratum into three zones of distinct post densities, from isotropic (the densest post array, zone 1) to most anisotropic (the sparsest post array, zone 3), and a zone of intermediate anisotropy (and post density, zone 2). We found that, although in all three zones cells displayed all three signaling patterns, the frequency of the cells displaying the RD pattern decreased and the frequency of the cells with PS pattern increased with decreasing post density (Supplementary Fig. 1A). The fraction of cells exhibiting the OS signaling dynamics remained approximately constant in all zones. Furthermore, when we used focal adhesion localization as a measure of cell polarity (Supplementary Fig. 2), we found that the fraction of cells polarized along the *y*-axis (in one or both directions), was approximately equal to the sum of the fractions of cells with PS and OS signaling patterns in each of the three substratum topography zones (Supplementary Fig. 1A). Cells with PS and OS signaling patterns also had a lower degree of spatial randomness of PI3K signaling and had more elongated shapes vs. the cells with the RD pattern (Supplementary Fig. 1B-D). Furthermore, we found a decrease in the average spatial randomness and increase in average cell elongation with decreasing post density, consistent with a higher occurrence of cells with the PS patterns (Supplementary Fig 1A-D). Interestingly, in spite of being polarized, the cells with the OS patterns displayed very low migratory persistence, similar to that of the cells with the RD signaling patterns (Supplementary Fig. 1E,F and 3). On the other hand, the migratory persistence of cells with the PS signaling patterns was greater than cells having the other two PI3K signaling patterns (Supplementary Fig. 1E,F and 3). These results were consistent with the hypothesis that the spatially localized PI3K activity is indeed enriched at the fronts of migrating cells. In particular, the limited migration of the cells with the OS signaling patterns is a reflection of continuous alteration in the direction of the front-rear polarity, with cells thus remaining polarized along the *y*-axis, but not persistently moving along it.

**A single mechanism can quantitatively account for different cell polarity and migration patterns**

What might account for distinct spatial PI3K signaling patterns and different migratory and polarization characteristics of 1205Lu cells on nano-patterned surfaces? Extracellular matrix components, including fibronectin (FN), can stimulate PI3K signaling (36, 37). An increasing engagement of ECM can also lead to stimulation of another member of the Rho family of small GTPases, RhoA (38, 39). We indeed found that increasing FN surface density stimulated, in a dose-dependent fashion, both PI3K activity (as evaluated by phosphorylation of its substrate, Akt) and RhoA activity (as evaluated by the activity of a RhoA-dependent kinase, ROCK, leading to phosphorylation of myosin light chain) (Supplementary Fig. 4). Furthermore, pharmacological inhibition of PI3K led to a decrease in Akt phosphorylation, but also to a proportional downregulation of a Rac1-dependent kinase PAK, further suggesting coordinated activation of Rac1 and PI3K signaling (Supplementary Fig. 5). Rac1 and PI3K activation can lead to actin polymerization and formation of protrusive lamellipodia (40, 41). We indeed found that localized PI3K was accompanied by local protrusive activity causing directional migration (Supplementary Fig. 6). Formation of lamellipodia can increase local ECM engagement, and thus further increase local ECM-dependent Rac1-PI3K activity leading to a putative positive feedback. However, lamellipodia extension can also increase cell-ECM contact and thus the local ECM-dependent activation of RhoA-ROCK and myosin-mediated contractile forces, which may trigger the lamellipodial retraction, constituting a negative feedback. The interplay between these spatially distributed mechano-chemical feedback interactions, as well as the more direct antagonistic interaction between Rac1 and RhoA (42, 43), can potentially lead to complex local signaling activity patterns. We thus mathematically explored if this postulated signaling network can account for the RD, OS and PS patterns of PI3K activity and the corresponding regulation of cell polarity and migration persistence.

Multiple feedback interactions outlined above as possibly controlling cell polarity can be described as several alternative hypotheses, which can be quantitatively expressed in a series of alternative mathematical models (full details in the companion modeling paper, Holmes *et al.* (44)). Their behavior qualitatively corresponds to the well-known dynamical system theory wherein a bistable system can be driven to oscillate by slow negative feedback (45). However, as explained in detail in the companion study, different models can generate distinct experimentally verifiable predictions that we have tested experimentally, arriving at a single mechanism (Fig. 2a) that proved to be consistent with all observations. In this mechanism, mutual antagonism of Rac1 and RhoA (correlated respectively with PI3K and ROCK signaling) gives rise to robust polarity, provided the total amount of each GTPase in the cell is sufficiently limited (a conservation constraint). As an added effect, we represented the mechanical coupling of lamellipodia (e.g. due to a build-up of membrane tension, (46, 47)) by assuming that lamellipodia mechanically compete, mutually inhibiting each other's extension. The Rac-Rho bistability was modulated by the positive (Rac1-dependent) and negative (RhoA-dependent) feedback to lamellipodia extension in an ECM-dependent fashion, as described above.

When described mathematically (see Supplementary Data for detail), this model (denoted the "Hybrid Model" in Holmes et al. (44)) successfully accounted for diverse patterns of localized Rac1-PI3K activity and lamellipodia-based local ECM engagement dynamics (Fig. 2b). In particular, in distinct parameter regimes, this mechanism generated either highly persistent Rac1-PI3K activation on one side of the cell and inhibition on the other side (consistent with the PS signaling pattern and persistent migration observed experimentally), or out of phase oscillatations of activity on both sides of the cell (consistent with the OS signaling pattern). Alternatively, the model predicted that the activities on both sides of the cell could be low, and thus potentially sensitive to stochastic fluctuations in signaling activity and/or ECM engagement. This behavior is interpreted as leading to random local activity patterns (consistent with the RD signaling pattern observed experimentally). These three patterns of model behavior can be conveniently represented in a bifurcation diagram, illustrating the dependence of cell dynamics on basal (i.e., intrinsic) Rac1 and RhoA activation rates (Supplementary Fig. 7).

To generate experimentally testable predictions for melanoma cells analyzed here, this mathematical model had to be first constrained ('trained') using experimental data to (a) define the ranges of model parameter values and (b) understand how the inherently 'noisy' cell populations with cell properties corresponding to distinct model parameter values could be 'mapped' onto the bifurcation diagram described above. Furthermore, it was also important to reflect in the model the local differences in the nano-topographic structure of the cell adhesion substratum. This, in turn, may limit, in a post spacing-dependent way, the ability of a cell to form extensive contact with the posts and thus the ECM-mediated input into the FA-triggered signaling pathways. This hypothesis was strongly supported by the analysis of the depth of FA locations, visualized by vinculin staining in our prior observations (35). Therefore, we assumed that the average ECM contact and thus 'intrinsic' Rac1 and RhoA-ECM, RhoA activation rates would depend on the local ECM-cell contact, and thus change from lowest in zone 1 (highest post density) to highest in zone 3 (lowest post density). Since we also observed considerable variability of cell penetration into the inter-post spacing, we further hypothesized that this variability would translate in each zone into the variability in the extent of cell-ECM contact and thus into corresponding variability in 'intrinsic' Rac1 and RhoA-ECM activation rates. We thus assumed that cell populations in each topographic zone could span a considerable range of parameter values and thus cross the boundaries separating different signaling regimes in the bifurcation diagram. To further constrain the extent of cell heterogeneity and the values of the model parameters, we had fit the model to the experimental results, as shown in Fig. 2d. This completed the 'model training' for melanoma cells cultured on topographically complex substrata, and allowed us to explore the predictive power of the parameterized model, and, thus the plausibility of the assumed underlying mechanism of signaling and migration patterns in this system.

**The proposed mechanism is validated by a range of perturbations of the postulated molecular network**

We next tested the postulated mechanism by performing a series of theoretical and experimental perturbations affecting assumed molecular interactions, as shown schematically in Fig. 2a. First, we perturbed the ECM input levels, which was experimentally achieved by varying the coating density of fibronectin, [FN]. We explored whether this would alter the relative frequencies of PI3K signaling patterns and the corresponding polarity and migration phenotypes. The critical model prediction was that this perturbation would shift a single boundary of the bifurcation diagram, separating the PS and OS signaling regimes. Thus, the model predicted that the fraction of cells displaying the RD signaling pattern would not change as ECM density was varied, while the frequency of the cells displaying the PS would decrease with increasing ECM (Figs. 2c,d). Keeping model parameters the same as determined during model training for 10 μg/ml [FN] condition, we extended the model to quantitatively predict cell behavior in the new 2 and 50 μg/ml [FN] condition (see Supplementary Data for detail), determining the corresponding deviations in two model parameters characterizing changes in the ECM-induced RhoA and Rac1 activities (Supplementary Figs. 4,5). Using this simple model adjustment, the model was further used to predict frequencies for all PI3K patterns for all topographic zones, with model predictions closely matching the experimental findings (Fig. 2c,d and Supplementary Fig. 8). Consistent with model predictions (assuming direct correspondence between the PS pattern of PI3K activity and the extent of cell migration), we found that cells moved less persistently with increasing [FN], and that the persistence increased at all [FN] values with decreasing post density (See Fig. 2e, and note that the ratio of net displacement, $D_y$ to total distance, $T_y$ increases, consistent with increased persistence).

We next perturbed another node of the signaling system (Fig. 2a): modulating the influence of Rac1 activity on lamellipodium formation and thus ECM input controlled by the strength of the local Rac1-mediated signaling (48, 49), with 10 μM of Rac1 inhibitor, NSC23766. The mathematical model indicated that, following this perturbation, the fraction of cells with the RD pattern would not change. However, this perturbation was predicted to increase the fraction of PS cells, while decreasing the fraction of OS cells (Fig 3a). We indeed found that this prediction was confirmed by experiments with individual cells frequently changing their signaling patterns from OS to PS soon after addition of the Rac1 inhibitor (Supplementary Figs. 9A,B and Supplementary movie 4). We quantitatively modeled the effect of Rac1 inhibition by adjusting a single model parameter, the ECM-mediated Rac1 activation rate (ECM-Rac1 activation rate), to match a single experimental data point (see Supplementary Data in detail). As before, this allowed us to predict the rest of the experimental results, with a very good match between the model and experiments (Fig. 3b). Again, as predicted by the model, the persistence of cell migration was enhanced by Rac1 inhibition in all post density zones (Fig. 3e).

We then performed similar experimental and theoretical perturbation of another parameter: the dependence of ROCK activity on ECM. Surprisingly, given that ROCK displays activity antagonistic to Rac1, inhibition of

ROCK activity with 10 μM of ROCK inhibitor, Y27632 (Fig. 3c) was predicted to have similar effects to that of inhibition of Rac1, i.e., increasing the frequency of PS cells. We indeed experimentally observed an increase in persistent signaling and cell migration following this perturbation (Fig. 3e, Supplementary Figs. 9C,D and Supplementary movie 5), with model predictions again yielding excellent quantitative agreement with experimental results. In this vein, we tested the predictions of the model regarding the dependence of PI3K signaling pattern on ROCK activity with a distinct, independent perturbation. Recently, the inhibition of microtubules has been reported to promote oscillation of cell protrusion (20). Treatment with 100 nM of nocodazole interferes with the polymerization of microtubules, preventing GEF-H1, a RhoGEF from being sequestered on microtubules, which causes GEF-H1 to be released into cytoplasm where it eventually activates ROCK(50). The model predicts that this perturbation would have an opposite effect to that of inhibition of ROCK (Fig. 4a). Indeed, we confirmed that the inhibition of microtubule polymerization activated GEF-H1 and subsequently ROCK although it did not change PI3K activity (Fig. 4b and Supplementary Fig. 10A). Furthermore, we found a reduction of persistent PI3K signaling and cell migration experimentally, the opposite observation to that of ROCK inhibition (Figs 4c,d, Supplementary Fig. 10B,C and Supplementary movie 6).

**The proposed mechanism links a common mutation to cancer cell invasiveness**

Finally, we investigated whether the model was capable of predicting the behavior of a less invasive melanoma cell line, SBcl2 (Fig 5). The invasive cell line, 1205Lu, used in our study thus far, carries a well-established genetic alteration, loss of expression of PTEN, a phosphatase antagonizing PI3K signalling (24), whereas in SBcl2 expression of PTEN is unperturbed. Consistent with this change, we found that, compared to 1205Lu cells, SBcl2 cells displayed lower PI3K activity and higher ROCK activity (Fig. 5a). When incorporated into the model, these perturbations of the signaling network found in non-invasive melanoma cell line resulted in the prediction that both bifurcation diagram boundaries (separating PS from OS, and OS from RD signaling patterns) would be shifted, thus increasing the fraction of cells with the RD pattern, primarily at the expense of cells with the PS signaling pattern (Fig. 5b). These predictions were supported by experimental observations (Fig. 5c). Furthermore, again consistent with the analysis, we found that both cell polarity and the persistence of cell migration decreased in SBcl2 cells relative to 1205Lu cells (Fig 6c-e), suggesting that more invasive cells can undergo a more directional migration while guided by chemical and mechanical cues provided by ECM.

**DISCUSSION**

The results presented here argue that diverse patterns of signaling and migratory cell responses guided by ECM organization, as well as genetic alterations or pharmacological perturbations of signaling components within a cell, can be explained by a simple mechanism supported by a quantitative mathematical model. The integrative, mathematical-experimental analysis leading to these results can provide a powerful tool for understanding normal and pathological phenomena associated with cell migration. This mechanistic understanding, supported by extensive experimental validation, strongly suggests the importance of mechano-chemical regulatory network, regulating lamellipodial extension through both positive (via Rac1) and negative (via RhoA) feedback stemming from modulation of the lamellipodia-ECM interactions. Conservation of the total amount of small GTPases in a cell effectively leads to a double negative feedback between the front and the back of the cell, preventing simultaneous high activation in both front and back cellular compartments. This simple mechanism and the associated model provide a powerful way to translate biochemical and genetic intracellular interactions and the structural and biochemical complexity of the surrounding ECM into a single analytical framework. The bifurcation diagram resulting from this model provides an intuitive and quantitative explanation of cell locomotion, wherein a domain of instability corresponding to the oscillatory migration separates the domains of random and persistent migration patterns. Our results show that genetic and pharmacological perturbations can shift the borders separating these domains, making specific migration patterns more likely to occur. Specific cell populations can be 'mapped' onto this diagram providing precise quantitative predictions of fractions of cells undergoing distinct migration responses. Therefore, heterogeneity of cell properties, or cellular 'noise', commonly observed in cell populations can allow the same population to explore diverse phenotypic responses. The 'mapping' can be unique to specific cell types, thus producing diverse results, with the underlying 'phenotypic diagram' universally accounting for the chemical nature of the signaling network.

It is important to assess the predictive power of the model developed here beyond the experimental system under investigation, i.e., the potential universality of the proposed mechanism. Previous analysis indicated that depletion of zyxin, a protein mediating coupling between ECM and intracellular signaling components, induced oscillatory migration patterns in fibrosarcoma cells (21). This result is consistent and can explained by our analysis, which underscores interaction between ECM and small Rho family GTPases as critical for determining the relative fraction of oscillating cells. Furthermore, our model explains why inhibition of ROCK in neutrophil-like cells increases the propensity for persistent cell migration, whereas microtubule disruption decreases it (51), and why microtubule disruption promotes oscillatory migration in retinal pigment epithelial cells (20). The model also provides an explanation for the prior findings that inhibition of Rac1 can increase persistence of cell migration (52). Combined with our study in melanoma cells, the predictive power of the model argues that the underlying mechanism may indeed represent a universal account of the emergence of diverse migration modes triggered by cell interaction with ECM.

Our results also shed light on an ostensibly paradoxical localization of both Rac1 and RhoA activities to the front of many migrating cells (53). Although, Rac1-PI3K activity is frequently associated with the cell 'frontness' and RhoA activity with the cell 'backness', our data indicate that RhoA can be activated if the ECM-plasma membrane contact at the cell front becomes sufficiently extensive, which might arrest the extension of a lamellipodium and lead to its retraction. This dynamic view of cell migration, with the possibility of re-polarization suggests that, locally, Rac1 and RhoA activities are in a continuous 'balancing act', with some degree of ECM-triggered activity for both of these small GTPases present at any time. This dynamic balance can endow a migrating cell with higher directional flexibility as observed in our analysis.

The mechanism of cell migration directionality supported by our analysis also suggests that the topography of the cell adhesion substratum defines the cell migration patterns in large part by controlling the extent of cell-substratum contact. Increasing cell-matrix contact in sparser topographies can enhance activation of extracellular matrix-regulated signaling pathways, which increases the migration persistence. The anisotropic nature of the substratum in these sparser migration zones provides an additional cue, specifying migration directionality. Previously, we provided experimental evidence for differential cell-matrix contact on these substrata, supporting this hypothesis (35). This mechanism is consistent with previous evidence emphasizing the importance of differential cell compliance with the nano-scale texture of adhesive surfaces (54, 55), suggesting that this mechanism of sensing of the nano-scale features can be very general.

The findings presented here also argue that mutations leading to a switch to more aggressive and invasive expansion in melanoma and other invasive cancers might increase the fraction of cells undergoing persistent migration, and thus following more faithfully the cues provided by the organization of the surrounding ECM. Synergic with ECM organization that can be highly anisotropic and well-defined in many normal tissues, such as the dermis underlying normal or cancerous melanocytes, or can become highly anisotropic during progression to invasive growth, as in invasive cancer progression (56, 57), a more persistent cell migration can enhance the spread of the invading cells guided by these non-random ECM cues. Furthermore, our prior analysis suggested that mutational changes observed in aggressive, invasive melanoma cells can direct them towards sparser matrix areas, which might further enhance cell-matrix contact and increase persistence of cell migration (35). These considerations can provide important insights into the still poorly understood biology of invasive cancer growth and emphasize the importance of accounting for the structure of the surrounding matrix in guiding this dangerous phase of cancer progression.

Overall, our results argue that, in spite of the complexity of biochemical and mechanical processes leading to cell locomotion in complex micro-environments, quantitative mechanisms can be proposed that can account for this complexity and accurately predict the effects of a diverse set of genetic, pharmacological and biochemical perturbations of cells and their milieu. These mechanisms are general, and thus applicable to many cell types, and yet quantitative and predictive enough to allow for precise description of diverse cell populations adopting a multitude of cell migration patterns. We anticipate that this and other examples of such quantitative modeling of signaling and phenotypic behavior will be essential for both fundamental description of cell biology, and for the development of more precise and targeted therapeutic interventions.

## MATERIALS AND METHODS

**Fabrication of Topographical Patterned Arrays.** We used polyurethane acrylate (PUA) as a mold material from a silicon master as previously described elsewhere (58), to construct graded post density array. Briefly, we drop-dispensed the ultraviolet (UV)-curable PUA onto a silicon master and then brought a flexible and transparent polyethylene terephthalate (PET) film into contact with the dropped PUA liquid. Then, the film was exposed to the UV light ($\lambda$= 200-400 nm) for 30 s through a transparent backplane (dose = 100 mJ cm$^{-2}$). After UV curing, we peeled off the mold from the master and further cured overnight. The resulting PUA mold used in the experiment was a thin sheet (~50 μm thick).

The topographic patterns with variable local density and anisotropy were fabricated on glass coverslips, using UV-assisted capillary molding techniques (35). Before applying the PUA mold onto the glass substratum, we cleaned the glass substratum with isopropyl alcohol, thoroughly rinsed in distilled ionized water, and then dried in a stream of nitrogen. Next, we spin-coated an adhesive agent (phosphoric acrylate: propylene glycol monomethyl ether acetate = 1:10, volume ratio) to make a thin layer (~100 nm) for 30 s at 3000 rpm. A small amount of the same PUA precursor was drop-dispensed on the substrate and a PUA mold was directly placed onto the surface. Capillary action allows the PUA precursor spontaneously to fill the cavity and then, this was cured by exposure to UV light ($\lambda$= 250-400 nm) for ~30 s through the transparent backplane (dose = 100 mJ cm$^{-2}$). After curing, we peeled off the mold from the substrate using sharp tweezers. The nanofabricated patterns used in this study, has the post spacing gradient along the *x* direction. The diameter of nano-posts was 600 nm with the varying post-to-post spacing from 300 nm [spacing ratio (spacing/diameter) of 0.5] to 4.2 μm (spacing ratio of 7) in the *x* direction, but constant post-to-post spacing of 600 nm (spacing ratio of 1) in the *y* direction.

**Cell culture.** Melanoma cell lines, 1205Lu and SBcl2 from the Wistar Institute collection were gifted from Dr. Rhoda Alani (Boston University). We also transfected the plasmids coding for 3-phosphoinotitide-specific Akt pleckstrin-homology (Akt-PH) domain fused with GFP in 1205Lu and SBcl2 using lipofectamin$^{TM}$ 2000 transfection reagents (Thermo Fisher, 11608027) as described in product's manual. GFP-Akt PH plasmids were gifted by Dr. Jin Zhang's lab (UCSD)(59). We cultured the cells in Dulbecco's modified Eagle's medium (Gibco) supplemented with 10% fetal bovine serum (Gibco), 50 U/ml penicillin/streptomycin (Invitrogen) and 90% humidity and split 1:4 after trypsinization every 2-3 days. We attached a glass coverslip with the topographical patterned substratum onto the bottom surface of the custom-made MatTek dish (P35G-20-C) after removing the original glass. Then, cells were re-seeded on the pattern pre-coated by fibronectin (Sigma-

Aldrich) for 3 hours and incubated overnight. To change the activities of small GTPases, 5 µM of Y27632 (Sigma-Aldrich), 10 µM of NSC23766 (Tocris) and 100 nM of Nocodazole (Acros Organics) were used.

**Western Blot Analysis.** We lysed cells in RIPA buffer (Thermo Scientific®) with protease/phosphatase inhibitor cocktail (Thermo Scientific®), then centrifuged at 12,000 x g at 4 °C for 15 min. Next, we collected the supernatant and measured its protein concentration using BCA assay kit (Pierce). Protein samples were subsequently diluted with sample buffer, heated at 70 °C for 10 minutes and 10 ~ 20 µg of proteins were used for Western blotting. We incubated protein-transferred membrane in 1:2000 diluted solution of myosin light chain (Abcam, ab11082), 1:1000 diluted solution of phosphorylated myosin light chain (Cell signaling, 3671), 1:2000 diluted solution of Akt (Cell signaling, 9272), 1:1000 diluted solution of phosphorylated Akt (Cell signaling, 9271), 1:1000 diluted solution of PAK (Cell signaling, 2602), 1:1000 diluted solution of phosphorylated PAK (Cell signaling, 2601), 1:1000 diluted solution of myosin IIa (Cell signaling, 3403) and 1:1000 diluted solution of phosphorylated myosin IIa (Cell signaling, 5026), β-actin (Santa Cruz, sc-1616) and 1:1000 diluted solution of GAPDH (Abcam, ab9484) in blocking solution at 4 °C overnight.

**Time-Lapse Microscopy.** For long-term observation, we mounted samples onto the stage of a motorized inverted microscope (Zeiss Axiovert 200M) equipped with a Cascade 512B II CCD camera that has an environmental chamber for maintaining 37°C and 5% $CO_2$. We used the Slidebook 4.1 (Intelligent Imaging Innovations, Denver, CO) for 3 hours at 10 min intervals or for 2 hours at 2 min intervals for automatic phase-contrast and epi-fluorescent live cell imaging.

**Tracking Cells and Quantification of Cell Shape/Migration.** With the help of the fluorescence-tagged Akt-PH in melanoma cells, we tracked cells automatically by thresholding epi-fluorescent images taken in a time-lapse fashion. Infrequently, we also manually segmented images that were poorly segmented in the automatic procedure, by outlining the boundary of each cell. Then, using MATLAB Image Processing Toolbox (The MathWorks, Natick, MA), the positions of cell centroids were calculated as a function of time and the persistence of cell migration, $D_y/T_y$, (the ratio of the shortest distance in y-axis direction between the initial location of an individual cell and its final location ($D_y$) to the length of whole trajectory of the cell during live cell imaging ($L_y$)), was measured. Axial ratio representing the degree of elongation of cells was defined as the ratio of length of longest axis of cells in y-axis to that in x-axis. Statistical analysis was performed using two-paired student t-tests.

**Quantification of spatial PI3K translocation.** Image segmentation from epi-fluorescent images was performed as follows. We read the pixel values representing the intensity in manually segmented single cells with MATLAB Image Processing Toolbox. Then, hot spots, the regions on the cell membrane having higher intensity value than the averaged intensity values of total pixels in a single cell, were further segmented out. We acquired the area ($A$), average intensity ($F$), and centroid coordinates of each hot spot region to calculate a signaling vector. The position of the hot spot $x_i = (x_i, y_i)$, relative to the cell centroid, was defined by subtracting the coordinates of the cell's centroid from those of the hot spots, $i$. The vector, $s_i$ had the magnitude proportional to the fluorescence intensity integrated over the overall area of the hot spots, $(A_iF_i)$, to account for the corresponding signaling intensity; and the overall signaling vector, $S$, describing the spatial PI3K activity polarization for a given cell was calculated as the sum of all $s_i$, as described in Fig. 1c (60) and equations below.

$$s_i = A_iF_i \frac{x_i}{\sqrt{x_i^2 + y_i^2}}; \qquad S = \sum_{N}^{i=1} s_i$$

For calculating the spatial randomness of a polarization signal, a virtual circle around the centroid of an individual cell was drawn and divided into 4 sectors evenly (upper and lower along the $y$-axis of the pattern, right and left along the $x$-axis). Then, the probabilities of the signaling vector localization to each sector over 3 hours of measurements ($p_i$) were determined and used for entropy ($E$) calculation. This entropy value ($E$) was used to evaluate spatial randomness according to the following equations (Supplementary Fig. 11).

$$E = -\sum_{i=1}^{4} p_i \log_2 p_i \qquad \text{Spatial randomness} = 2^E$$

On a more intuitive level, if the probabilities of the signaling vectors located in either upper or lower sectors along the $y$-axis of patterns were below 0.6, we considered this polarity pattern to be random. Otherwise, if the difference between the probabilities of the signaling vector localzing within either upper or lower sections vs. the opposite section were below 0.2, we categorized the cells as oscillatory. The rest of cases, i.e, when the differences of probabilities were over 0.2, were considered to display persistent polarization dynamics.

**Immunofluorescence Staining.** We re-plated cells on the nanofabricated coverslip after fibronectin coating as described above. Samples were fixed with ice-cold 4% paraformaldehyde for 20 minutes and permeablized with 0.1% Triton X-100 in PBS for 5 minutes. After washing with ice-cold PBS, we blocked samples with 10% goat-serum for 1 hour, and then incubated them with primary antibody against vinculin (1:200, Sigma-Aldrich, V9131) for 3 hours at room temperature. After subsequent washing, the samples were incubated with a secondary antibody for vinculin (1:500), with Alexa fluor 594 conjugated phalloidin (1:40, Molecular Probes) and/or Hoescht (Invitrogen) for 1 hour at room temperature. The slides were mounted with anti-fade reagent (SlowFade gold, Invitrogen) and imaged by confocal microscope (Zeiss 510 Meta confocal) with a x63 oil immersion objective (Zeiss, 1.6 NA).

**Quantification of polarized FA localization** As quantifying spatial PI3K dynamics, a virtual circle around the centroid of an individual cell was drawn and divided into 4 sectors evenly (upper and lower along the *y*-axis of the pattern, right and left along the *x*-axis). Then, the probabilities of the FA localization to each were determined and used for categorization. If the probabilities of the FA localization located in either upper or lower sectors along the y-axis were over 0.8, we considered this FA localization to be polarized. Otherwise, we categorized the cells having random FA spatial distribution.

**Active RhoGEF assay** Active Rho GTPase exchange factor (RhoGEF) pull-down assay has been described in detail elsewhere (61, 62). Briefly, cells were lysed in RhoGEF buffer, which is combined with 20mM HEPES, 150 mM NaCl, 5mM $MgCl_2$, 1% (vol/vol) Triton X-100, 1mM DTT and protease inhibitor cocktail. 15 g of GST – $RhoA^{G17A}$ beads were then added to each equalized lysates and samples were rotated at 20 rpm for 45 min at 4 ºC. After washing beads three times in the RhoGEF buffer, SDS-PAGE analysis was processed and performed immunoblotting with GEF-H1 antibody (Cell signaling, 4076S). $RhoA^{G17A}$ mutant plasmid was a generous gift from Dr. Rafael Garcia-Mata at the University of Toledo.

## FIGURE LEGENDS

**Figure 1. A melanoma cell population displays three coexistent polarity patterns whose relative abundance is modulated by the topography of the cell adhesion substratum** **a.** A scanning electron microscopy image of a 1205Lu (aggressive melanoma cell line) cell cultured on fibronectin (FN) pre-coated post arrays. **b.** Schematic of a single 1205Lu cell sensing anisotropic topographic cues from the nanofabricated graded density nanopost array substrata. In the dense post region (left), the nanopost density is equal in *x* and *y* directions, whereas, in the sparse post region, where penetration between posts of cells results in greater contact with the substratum adhesion, the substratum anisotropy is the greatest due to a much larger spacing in the *x* but not *y* direction. **c.** The angle of a signaling vector (θ) of an individual cell is defined as the direction from its centroid to the sum of centroids of its 'hot spots' of PI3K activity with respect to the *x* direction. Details are described in the Materials and Methods. **(d-f)** Three dynamic patterns of cell polarity seen in sequential images of PI3K activation and indicated by the signaling vector defined in (c) in three representative cells. Representative pseudo-color images of elevated spatial PI3K activation (color scale at the bottom) (left panels) and sinθ(t) at each time point (arrows) (right panels), with the indicated time resolution; (**d**) the random polarity pattern (**e**), the oscillatory polarity pattern and (**f**) and persistent polarity pattern.

**Figure 2. The proposed molecular mechanism and associated mathematical model can quantitatively account the observed dynamic polarity patterns. a.** Mechanism schematic (top) and classification of cell behavior (bottom). RhoA-ROCK signaling and Rac1-PI3K signaling are mutually inhibitory in coupled signaling networks activated at the front or rear of a cell. Signals from the extracellular matrix activate RhoA

(and possibly Rac1-PI3K, see the text). Rac1-PI3K induced cell spreading increases contact with (and signaling from) ECM. RhoA-ROCK induced cell contraction decreases the contact with (and signaling from) ECM. The model accounts for the three main classes of behaviors: Random (no Rac1-PI3K dominated "hot spots" established), oscillatory (cycling between hot spots in the front and rear), and persistent (polarization of hot spot to one lamellipodium). **b.** The mathematical model emerging from the mechanism described in panel (a) leads to the bifurcation diagram specifying parameter domains leading to each of 3 different polarity patterns (the bifurcation parameters are the intrinsic activation rates of GTPases, RhoA and Rac1, see the model description in the text). Outboxes at the top show the dynamics of a pseudopod-ECM contact at the front (Cell-ECM contact$_{front}$) and at back (Cell-ECM contact$_{back}$) of a model cell for random, oscillatory, and persistent polarity patterns. **c**. Cell populations cultured in different nano-post density zones are mapped onto the bifurcation diagrams as described in the text. Intensity of grey zones represents the density of posts in nanofabricated substrata (lightest grey represents densest post region). These regions are inferred based on experimentally determined fractions of cells with persistent polarity in zone 3 for each FN density condition. Variation of ECM (FN) density is predicted to shift the oscillatory/persistent bifurcation boundary thus changing fractions of cells adopting these patterns in a population. **d.** Comparison of modeling predictions (*in silico*) and experimental results (*in vitro*) for the dependence fractions of different cell polarity patterns on surfaces with varying post and FN densities. Dashed lines indicate the fraction of cells having polarized focal adhesion (FA) localization (See details in Supplementary Fig.2 and Material and Methods). **e.** Variation of migration persistence metric, $D_y/T_y$, correlates with the polarity patterns ($D_y$ is the distance travelled by the cell along the y direction, and $T_y$ is the total distance travelled by the cell along the y direction) over 3 hours for different post and FN densities. * = Statistical significance of $D_y/T_y$ among each group, *$P<0.05$ and all paired Student's t-test.

**Figure 3. Effect of pharmaceutical perturbations of Rho GTPase-dependent signaling on relative abundance of distinct polarity patterns.** Top: Blue arrow depicts Rac1 inhibition by Rac1 inhibitor, NSC23766 (**a**) and ROCK inhibition by ROCK inhibitor, Y27632 (**c**), which biases cells towards persistent dynamics (shown in the ECM-induced Rac1 and RhoA activation rates parameter plane). Bottom: Rac1 inhibition (**a**) and ROCK inhibition (**c**) shift the boundary between oscillatory and persistent regime to the left, increasing the persistent regime and hence the fraction of cells exhibiting persistent dynamics. Comparison of model (*in silico*) and experiment (*in vitro*) results for the dependence of PI3K dynamics on nanofabricated substrata coated with 50 μg/ml of FN after Rac1 inhibition (**b**) and ROCK inhibition (**d**). Dashed lines indicate the fraction of cells having polarized FA localization on corresponding experimental conditions**.** Dependence of the persistence metric, $D_y/T_y$ evaluated over 3 hours on the post density with Rac1 and ROCK inhibition. Both inhibitions enhance the persistence of cell migration. *$P<0.05$ for $D_y/T_y$ values in each group using all paired Student's t-test.

**Figure 4. Effect of pharmacological perturbation of microtubule dynamics on dynamic cell polarity patterns. a.** Left: Blue arrow depicts increasing RhoA-ROCK activity by microtubule inhibition, which biases cells towards oscillatory dynamics. Right: microtubule inhibition shifts the boundary between oscillatory and persistent regime to the right decreasing the persistent regime and hence the fraction of cells exhibiting persistent dynamics. **b.** Microtubule inhibition increases RhoA-ROCK activity as measured by immunoblotting of phosphorylated Myosin IIa/Myosin IIa and phosphorylated MLC/MLC, which potentially results from higher activation of GEF-H1, a RhoGEF whose activity is affected by nocodazole (see RhoGEF activity assay described in Materials and Methods). **c.** Comparison of model (*in silico*) and experiment (*in vitro*) results for the dependence of PI3K dynamics on post density with 10 μg/ml of FN after microtubule inhibition. **d.** Dependence of the ratio, $D_y/T_y$ for 1 hour on post density substrata and nocodazole: Inhibition of microtubule reduces the persistence of cell migration. *$P<0.05$ for $D_y/T_y$ values in each group using all paired Student's t-test.

**Figure 5. The alteration of dynamic polarity patterns in metastatic cells vs. non-invasive melanoma cells is accounted by the proposed feedback mechanism. a.** SBcl2 cells display lower Rac1-PI3K activity and higher RhoA-ROCK activity than 1205Lu cells, as measured by immunoblotting of phosphorylated Akt and Akt (Rac1-PI3K) and phosphorylated MLC and MLC (RhoA-ROCK). **b.** Assuming a decrease in ECM-Rac1 activation rate and increase in ECM-RhoA-ROCK activation rate in SBcl2 cells shifts the boundary between oscillatory and persistent regime and boundaries between random and oscillatory regimes towards the right, predicting smaller fraction of cells in the persistent Rac1-PI3K regime and a larger fraction in the random regime. **c.** Comparison of model (*in silico*) and experiment (*in vitro*) results for the dependence of PI3K dynamics of SBcl2 cells on nanofabricated substrata coated with 10 μg/ml FN. **d.** Axial ratio (the degree of aligned elongation) in the two cell lines. **e.** $D_y/T_y$ for 3 hours of the two cell lines. The noninvasive SBcl2 cells have inhibited elongation and directional cell migration. *$P<0.05$ for $D_y/T_y$ values in each group using all paired Student's t-test.

**Supplementary Figure 1. Cell elongation and migration regulated by PI3K dynamics depends on local adhesion substratum topography. A.** Fraction of cells showing random, oscillatory and persistent PI3K dynamics versus post density. Dashed lines indicate the fraction of cells having polarized FA localization. **B.** Spatial randomness of PI3K activity increases with post density. Each dot indicates the value for an individual cell and its color corresponds to the categorized PI3K dynamics. Bold lines are averaged spatial randomness of cells in each zone, described in Materials and Methods and Supplementary Figure 11, for cells within each zone. *$P<0.05$ for the comparisons of spatial randomness using all paired Student's t-test. (**C-F**) Dependence of the axial ratio (**C**) and the persistence metric, $D_y/T_y$ (**E**) on the post density. Cells with random (RD) polarity patterns are less elongated (**D**) and have higher random migration (**F**) than cells with oscillatory (OS) and

persistent (PS) polarity patterns. * = Statistical comparison of the persistence metric, $D_y/T_y$ and spatial randomness among groups, *P<0.05, **P<0.01, ***P<0.005 and all paired Student's t-test.

**Supplementary Figure 2. Dependence of the spatial distribution of focal adhesions (FAs) on the local post density A.** Classification of the spatial distribution of FAs as Random, (evenly distributed around cell periphery) and Polarized, (concentrated at the two cell ends). **B.** Cells displaying polarized distribution of FAs are more elongated vs. cells with random FAs distribution. * = Statistical analysis of th axial ratio, ***P<0.005 and all paired Student's t-test. **C.** Fractions of cells exhibiting random or polarized FA distribution as a fuctnion of the post density.

**Supplementary Figure 3. Persistence of cell migration vs. polarity factor**. PI3K dynamics was used to determine cell migratory dynamics and polarity patterns: RD cells exhibit random movement and OS cells oscillate in their position without overall advancement. PS cells migrate in one of the available *y*-axis directions in a highly persistent fashion.

**Supplementary Figure 4. ECM-stimulated ROCK and PI3K activity.** Increase in [FN] density enhances ROCK and PI3K activity. This data underlies the assumption in the model that higher ECM signaling increases ECM-Rac1 and ECM-RhoA and interaction, and thus induced Rac1 and RhoA activation rates.

**Supplementary Figure 5. Correlation between PI3K activity and Rac1 activity.** Dose-dependent effect of LY294002, a PI3K inhibitor, on Rac1-induced PAK phosphorylation and PI3K-induced Akt phosphorylation.

**Supplementary Figure 6. Correlation between PI3K dynamics and the direction of cell protrusion. A**. An illustration describing the angle of the signaling vector (θ) and the direction of cell protrusion (δ), i.e. the direction from a cell centroid at the initial time point (t) and that at the next time point (t+1). **B.** Positive correlation between θ and δ uncovered in the experimental measurements.

**Supplementary Figure 7. The prediction of PI3K signaling patterns for various combinations of the intrinsic Rac1 and RhoA activation rates A.** Bifurcation diagram of PI3K signaling patterns depending on the combinations of parameter values representing intrinsic Rac1 and RhoA-ECM activation rate **B.** Lamellipodia-based local ECM engagement dynamics (Cell-ECM) corresponding to the combination of parameter values indicated by same color dots as in panel A.

**Supplementary Figure 8. Comparison between *in silico* predictions and *in vitro* results for variable conditions: (A) the ECM density, (B) inhibition of ROCK, Rac1 and microtubule polymerization and (C) different genetic makeup in a non-invasive melanoma cells, SBcl2 cultured on FN (50 or 10 μg/ml) - coated substrata.** Lines are simulated boundaries between random and oscillatory polarity regimes and between oscillatory and persistent regimes. Dots and crosses correspond to calculated relative pattern

abundances based on the number of cells showing each type of PI3K dynamics in corresponding experimental results. Arrows indicate data used for model training for each experimental condition.

**Supplementary Figure 9. Dynamical change of spatial activity of Rac1-PI3K by perturbations of small GTPases of the Rho family A.** A pseudo-colored image showing time dependent spatial PI3K activity in a representative cell. A Rac1 inhibitor, NSC23766, applied at 45 min, abruptly shifts oscillatory polarity pattern to persistent polarization. (Color scheme and arrows are as in Fig 1). **B.** Time-course of the changes of sin (θ(t)), defined as in Fig 1, before and after application of NSC23766. **C.** A pseudo-colored image showing time dependent spatial PI3K activity in a representative cell. Y27632, applied at 45 min, abruptly shifts the oscillatory polarity pattern to persistent polarization. **D.** Time-course of the changes of sin(θ(t)) before and after application of Y27632.

**Supplementary Figure 10. Dynamical change of spatial activity of Rac1-PI3K in response to inhibition of microtubule polymerization A.** PI3K activity independent on microtubule inhibition measured by immunoblotting with Akt and phosphorylated Akt antibodies. **B.** A pseudo-colored image showing time dependent spatial PI3K activity in a representative cell. Nocodazole, applied at 30 min, abruptly shifts persistent polarity pattern to oscillatory polarization **C.** Time-course of the changes of sin(θ(t)) before and after application of nocodazole.

**Supplementary Figure 11. Description of signaling vector and spatial randomness.**

**Supplementary Figure 12. Raw western blotting images**

**SUPPLEMETNARY TEXT**

**Introduction**

As described in the main text, we model the regulation of cell spreading and contraction by small GTPases such as Rac1 and RhoA, with additional ECM input due to spreading. The axis of post density arrays permits a simplification to 1D, but we simplify even further, to consider a "two-lamellipod" model. Our model consists of a set of differential equations for GTPase activity at each of the two opposite lamellipods, accounting for feedback between Rac1, RhoA and ECM that cells contact locally across the cell. Emergent behavior includes bi-stability and oscillatory dynamics driven by slow negative feedback as described in Holmes *et al.* (44). We implemented a sequence of possible models, with distinct assumptions about the interactions and with distinct sub-models responsible for bi-stability versus negative feedback, but only one had qualitative properties in concordance with experimental observations. Here we use that "ultimate" model to investigate the correspondence between cells responding to variable degrees of anisotropic substrate input (due to post density of nanofabricated substrata) and the modification of environmental cues rewiring signaling the cells use for adjusting their behavior. Importantly, we explain how the model was "trained" and parameterized by a subset of the data and then validated by distinct data.

## 1. One-dimensional model cells

We assume that the key small GTPase-related activities controlling protrusion, contraction, and polarization, are Rac1-PI3K and RhoA-ROCK, and that the total amount of each GTPase is constant in the cell (conservation). For instance, local activation of Rac1-PI3K, which leads to cell protrusion and migration in a specific direction, depletes Rac1 at the opposite side of the cell, reducing the availability of Rac1 for activation. Likewise, spreading at one end of the cell creates additional cell-ECM contact, which both influences signaling locally, and limits spreading elsewhere. These features lead to competition between lamellipods where local cell protrusion at a "front" causes retraction at the opposite cell end, which becomes the "back". As shown here and in Holmes *et al* (44), this model cells can either migrate persistently or, under some conditions, oscillates via slow negative feedback of the bistable system creating a hysteresis loop.

## 2. Crosstalk between small GTPases

First, we modelled the mutual inhibition between Rac1-PI3K and RhoA-ROCK signaling with the simplest "toggle switch" equations:

$$\frac{dR}{dt} = b_R \frac{R_I}{1+\rho^n} - \delta R$$

$$\frac{d\rho}{dt} = b_\rho \frac{\rho_I}{1+R^n} - \delta \rho$$

where $R$ represents Rac1-PI3K activity (e.g. amount of active Rac1) and $R_I$ the amount of inactive Rac1 (available to be activated); similarly $\rho$, $\rho_I$ reflect the RhoA-ROCK activity (amounts of active and inactive Rho-A), $b_R$ and $b_\rho$ are intrinsic activation rates of Rac1 and RhoA, $\delta$ is the inactivation rate of small GTPases (assumed to be the same for Rac1 and RhoA). For convenience, and to reduce the number of free parameters, the GTPase concentrations have been scaled by the "$IC_{50}$" parameters of their mutual inhibition, such that $R=1$ and $\rho=1$ in the model lead to a 50% drop in the activation rate of each GTPase. This simple model is known to have emergent behavior that includes a bi-stable switch (i.e, high Rac1-PI3K and low RhoA-ROCK vs. Low Rac1-PI3K and high RhoA-ROCK) as well as a mixed state and states in which Rac or Rho dominate.

We implement the above two equations in each of the two "lamellipods" (1 and 2), and assume that, on the timescale of interest, there is conservation of the total amount of GTPase in the cell, i.e. $R_T = R_1 + R_2 + R_I =$ constant and similarly $\rho_T = \rho_1 + \rho_2 + \rho_I$.

$$\frac{dR_1}{dt} = b_R \frac{1}{1+\rho_1^n}(R_T - R_1 - R_2) - \delta R_1$$

$$\frac{dR_2}{dt} = b_R \frac{1}{1+\rho_2^n}(R_T - R_1 - R_2) - \delta R_2$$

$$\frac{d\rho_1}{dt} = b_\rho \frac{1}{1+R_1^n}(\rho_T - \rho_1 - \rho_2) - \delta\rho_1$$

$$\frac{d\rho_2}{dt} = b_\rho \frac{1}{1+R_2^n}(\rho_T - \rho_1 - \rho_2) - \delta\rho_1$$

## 3. Incorporation of Cell-ECM contact into the models

Rac1-PI3K signaling stimulates cell protrusion and results in increased contact between the cell and ECM. RhoA-ROCK signaling leads to contraction, decreasing that contact area. These affect the cell-ECM contact dynamics, $E_1$ or $E_2$ as described through the following equations.

$$\frac{\partial E_1}{\partial t} = \varepsilon(P(R_1) - C(\rho_1)k(E_1, E_2))$$

$$\frac{\partial E_2}{\partial t} = \varepsilon(P(R_2) - C(\rho_2)k(E_1, E_2))$$

where

$$P(R_i) = k_R + \gamma_R \frac{R_i^m}{R_0^m + R_i^m} \quad (i = 1 \text{ or } 2)$$

$$C(R_i) = k_\rho + \gamma_\rho \frac{\rho_i^m}{\rho_0^m + \rho_i^m} \quad (i = 1 \text{ or } 2)$$

$$k(E_i, E_j) = E_i(k_s E_i + \square_d E_j) \quad (i = 1, j = 2 \text{ or } i = 1, j = 2)$$

Here, $\varepsilon$ is a rate constant for cell spreading. We assume $\varepsilon < 1$ so that ECM- mediated negative feedback is slow compared to GTPase dynamics responsible for polarity. This is consistent with the idea that GTPase dynamics are faster than protrusion and contraction. $P(R_i)$ describes the Rac1-PI3K induced protrusion, $C(R_i)$ represents the RhoA-ROCK induced contraction and $k(E_i, E_j)$ describes competition between lamellipodia. $k_R$ corresponds to basal protrusion rate and $k_\rho$ is the basal contraction rate. $\gamma_R$ and $\gamma_\rho$ represent the magnitudes of Rac1 and RhoA mediated effects on lamellipodial protrusion and contraction rates, respectively, resulting in ECM signaling increase and decrease. Additionally, we assumed that cell-ECM contact activates RhoA-ROCK as described by the following form for $b_\rho$ (35, 53)

$$b_\rho = k_E + \gamma_E \frac{E_i^m}{E_0^m + E_i^m} \quad (i = 1 \text{ or } 2)$$

Here, $b_\rho$ is the RhoA activation rate, with $k_E$ the intrinsic RhoA activation rate and $\gamma_E$ the magnitude of ECM-induced RhoA activation rate (RhoA-ECM activation rate). Figure 2b shows the $b_R$ - $\gamma_E$ parameter plane (Rac1 activation rate vs RhoA-ECM activation rate) indicating the distinct regimes of PI3K dynamics. For parameter values, see Table 1.

## 4. Training parameters to explain the variable population of cells showing specific PI3K dynamics depending on the degree of post densities.

To train this model using the experimental data, we assumed the following

1) Cells are inherently 'heterogeneous' and the degree of heterogeneity depends on post densities that affect the extent of vertical pseudopod protrusion.

2) The substrate input that influences the extent of vertical pseudopod protrusion affects intrinsic Rac1 activation rate, $b_R$ and RhoA-ECM activation rate, $\gamma_E$.

Melanoma cells spread lamellipodia vertically as well as horizontally on post arrays. The vertical protrusion allows cells to receive additional substrate ("ECM signaling") input whose degree is determined by how far the lamellipodia can infiltrate between posts. On posts whose density is sparser, cells more easily infiltrate between posts. This conclusion is strongly supported by our prior observations (35). Intrinsic heterogeneity generates the variance of substrate input that we here associate with a range of Rac1 activation rate, $b_R$ and ECM-induced Rho activation rate, $\gamma_E$. Thus, we applied the following equations to describe the range of vertical protrusion from $P_{min}$ (the least protrusion) to $P_{max}$ (the greatest protrusion).

$$P_{min} = k_{p,min} d + s_{min}$$
$$P_{max} = k_{p,max} d + s_{max}$$

Here $k_{p,min}$ and $k_{p,max}$ are conversion coefficients ($k_{p,min} < k_{p,max}$) and $d$ is a *dimensionless distance* from the location of the cells to the point where the post density is greatest on the fabricated substratum. That is, we associate $d=0$ with cells located in the densest region of the post array with $d$ increasing with post spacing (e.g., larger $d$ is associated with sparser posts). To validate our model, we matched the ranges of $d$ to that of regions on the substrata having variable post densities ranging from 0.5 to 3.5: dense (the least penetrable zone, $0.5 \leq d < 1.5$), intermediate ($1.5 \leq d < 2.5$), and sparse zone (the most penetrable zone, $2.5 \leq d \leq 3.5$). These are normalized, scaled distances. $s_{min}$ and $s_{max}$ represent the minimum and maximum protrusion on the highest post density arrays ($s_{min} < s_{max}$). Additionally, we incorporated the assumption that $b_R$ and $\gamma_E$ are linearly dependent on the vertical protrusion, $P$ as below.

$$b_R = C_R P + w_R \quad (P_{min} < P < P_{max})$$
$$\gamma_E = C_\rho P + w_\rho \quad (P_{min} < P < P_{max})$$
$$b_{R,min} = C_R P_{min} + w_R = C_R(k_{p,min}d + s_{min}) + w_R = C'_{R,min} d + w'_{R,min}$$
$$b_{R,max} = C_R P_{max} + w_R = C_R(k_{p,max}d + s_{max}) + w_R = C'_{R,max} d + w'_{R,max}$$
$$\gamma_{E,min} = C_\rho P_{min} + w_\rho = C_\rho(k_{p,min}d + s_{min}) + w_\rho = C'_{\rho,min} d + w'_{\rho,min}$$
$$\gamma_{E,max} = C_\rho P_{max} + w_\rho = C_\rho(k_{p,max}d + s_{max}) + w_\rho = C'_{\rho,max} d + w'_{\rho,max}$$

where $C_R'$ and $C_\rho'$ are conversion coefficient from the dimensionless distance implying the post density where cells locate determining the degree of protrusion to corresponding intrinsic activation rates of Rac1 and RhoA. $w_R'$ and $w_\rho'$ are corresponding conversion constants from the dimensionless distance of cells from the densest post array, $d$, to intrinsic activation rates. These equations define the range of $b_R$ and $\gamma_E$ values resulting from varying levels of protrusion in different regions of the post array as $d$ varies from 0.5 (the densest region) to 3.5 (the sparsest region). To constrain conversion coefficients, $C_R'$, $C_\rho'$, $w_R'$ and $w_\rho'$, we use experimental results for the fraction of 1205Lu cells exhibiting different dynamics on the different regions of the substrate having variable post density. In the model, these different sections of the substrate and variation in cell properties in each section can be mapped onto variations of $b_R$ and $\gamma_E$ on a bifurcation diagram as shown in Figure 2c. Each grey stripe in Figure 2c outlines the range of $b_R$ and $\gamma_E$ corresponding to the sub-regions of different post density on the substrata. Each strip is divided into three regions by the bifurcation curves that delineate the boundary between different dynamic regimes (oscillatory, persistent, or random) of the model. The fraction of each stripe contained within a particular regime (the oscillatory regime for example) represents the fraction of cells (in that post zone) predicted by the model to exhibit that behavior. To train the model parameters, we first fixed values (taken from (Holmes et al. 2016)) for all parameters except ($b_R$ and $\gamma_E$) that ensure the model exhibits all three dynamic regimes. Subsequently, to determine the values of ($C'$ and $w'$) that describe the link between the degree of substratum anisotropy and Rac1-PI3K dynamics, we fitted the area of sectioned stripes to the fraction of cells showing each PI3K dynamic (Fig. 2d). For parameter values, see Table 2. We next discuss how experimental perturbations are represented in the model as changes of this base parameter set.

**5. Modification of parameters depending on pre-coated ECM density on the nano-fabricated substrata.**

We further evaluated our model with fitted parameters above as applying to variable pre-coated ECM density on the nano-fabricated substrata. It is widely accepted that the increase in ECM density enhances small

GTPases, both Rac1 and RhoA and Supplementary fig 5 supports this as well. To explain this relationship between ECM density and activities of Rac1 and RhoA, we assumed that ECM-Rac1 activation rate, $\gamma_R$ and ECM-RhoA activation rate, $\gamma_\rho$ are positively linearly-related to ECM density, [ECM] as below.

$$\gamma_R = a_R [ECM] + t_R$$
$$\gamma_\rho = a_\rho [ECM] + t_\rho$$

With this assumption and the conversion coefficients ($a_R$, $a_\rho$, $t_R$, and $t_\rho$) that we fitted from the experiments on 10 µg/ml of FN, except $\gamma_R$ and $\gamma_\rho$ (re-denoted here as $\gamma_{R\,[10]}$ and $\gamma_{\rho\,[10]}$), we found $\gamma_{R\,[50]}$ and $\gamma_{\rho\,[50]}$ well-matched to the population ratio of persistent dynamics within sparse zone on 50 µg/ml of FN and confirmed that $\gamma_{R\,[50]}$ and $\gamma_{\rho\,[50]}$ were larger than $\gamma_{R\,[10]}$ and $\gamma_{\rho\,[10]}$. In addition, for validation of newly changed parameters $\gamma_{R\,[50]}$ and $\gamma_{\rho\,[50]}$, we compared the rest of population ratios of persistent, oscillatory and random dynamics within each zone and confirmed well-fitted with experimental data on 50 µg/ml of FN. Furthermore, we extrapolated $\gamma_{R\,[2]}$ and $\gamma_{\rho\,[2]}$ from the equations describing the relationship between [ECM] and $\gamma_R$ and $\gamma_\rho$ and the fitted values of $\gamma_{R\,[10]}$ / $\gamma_{\rho\,[10]}$ and $\gamma_{R\,[50]}$ / $\gamma_{\rho\,[50]}$. The application of extrapolated $\gamma_{R\,[2]}$ and $\gamma_{\rho\,[2]}$ allowed us to validate our model with the experimental results on 2 µg/ml of FN as well. For parameter values, see Table 2 and Table 3

## 6. Modification of parameters depending on the perturbation of Rac1 or RhoA related signaling.

We adjusted $\gamma_R$ or $\gamma_\rho$ to selected experimental data for training, the population ratio of persistent dynamics within sparse zone, and tried to simulate the perturbation of Rac1 or RhoA-related signaling with the values. These adjusted values of $\gamma_R$ or $\gamma_\rho$ are consistent with the predictions, i.e, the decrease in $\gamma_R$ describing the inhibition of Rac1, the decrease in $\gamma_\rho$ describing the inhibition of ROCK, and the increase in $\gamma_\rho$ describing the inhibition of microtubule polymerization. Also, the rest of data other than the data used in the training were well matched with the predictions from the model applying the adjusted $\gamma_R$ or $\gamma_\rho$ without additional parameter fitting in all experimental data, inhibition of Rac1, ROCK and microtubule polymerization (Supplementary Fig 8). For parameter values, see Table 3.

| Parameter | Value |
|---|---|
| $R_T$ | 2 |
| $\rho_T$ | 2 |
| $\delta$ | 1 |
| $k_R$ | 1.25 |
| $k_\rho$ | 0.5 |
| $R_0$ | 0.75 |
| $\rho_0$ | 0.75 |
| $k_s$ | 0.1 |
| $l_d$ | 1 |
| $\varepsilon$ | 0.1 |
| $k_E$ | 5 |
| $E_0$ | 1 |
| $m$ | 3 |
| $n$ | 3 |

**Table 1. Parameters applied to simulation**

| Parameter | Value |
|---|---|
| $C'_{R,max}$ | 0.6 |
| $C'_{\rho2,max}$ | 0.0493 |
| $C'_{R,min}$ | 0.05 |
| $C'_{\rho,min}$ | 0.0478 |
| $w'_{R,max}$ | 2.4 |
| $w'_{\rho,max}$ | 0.1907 |
| $w'_{R,min}$ | 1.35 |
| $w'_{\rho,min}$ | 0.0403 |
| $a_R$ | 0.0008 |
| $a_\rho$ | 0.008 |
| $t_R$ | 0.71 |
| $t_\rho$ | 0.26 |

**Table 2. Parameters applied to mapping intrinsic Rac1 activation rate ($b_R$) and RhoA-ECM activation rate ($\gamma_E$) on nanofabricated substrata**

| Parameter | $\gamma_R$ | $\gamma_\rho$ | ECM condition | Figure |
|---|---|---|---|---|
| [FN]: 2 µg/ml | 0.712 | 0.28 | | 2 |
| [FN]: 10 µg/ml | 0.718 | 0.34 | | 2 |
| [FN]: 50 µg/ml | 0.75 | 0.66 | | 2 |
| Inhibition of Rac1 | 0.01 | 0.66 | [FN]: 50 µg/ml | 3 |
| Inhibition of ROCK | 0.75 | 0.132 | [FN]: 50 µg/ml | 3 |
| Inhibition of Microtubule | 0.718 | 0.65 | [FN]: 10 µg/ml | 4 |
| SBcl2 | 0.375 | 5 | [FN]: 10 µg/ml | 5 |

**Table 3 ECM-Rac1 activation rate ($\gamma_R$) and ECM-RhoA activation rate ($\gamma_\rho$) regulated by controlled environmental cues**

**REFERENCE**

1. Lauffenburger DA & Horwitz AF (1996) Cell migration: a physically integrated molecular process. *Cell* 84(3):359-369.

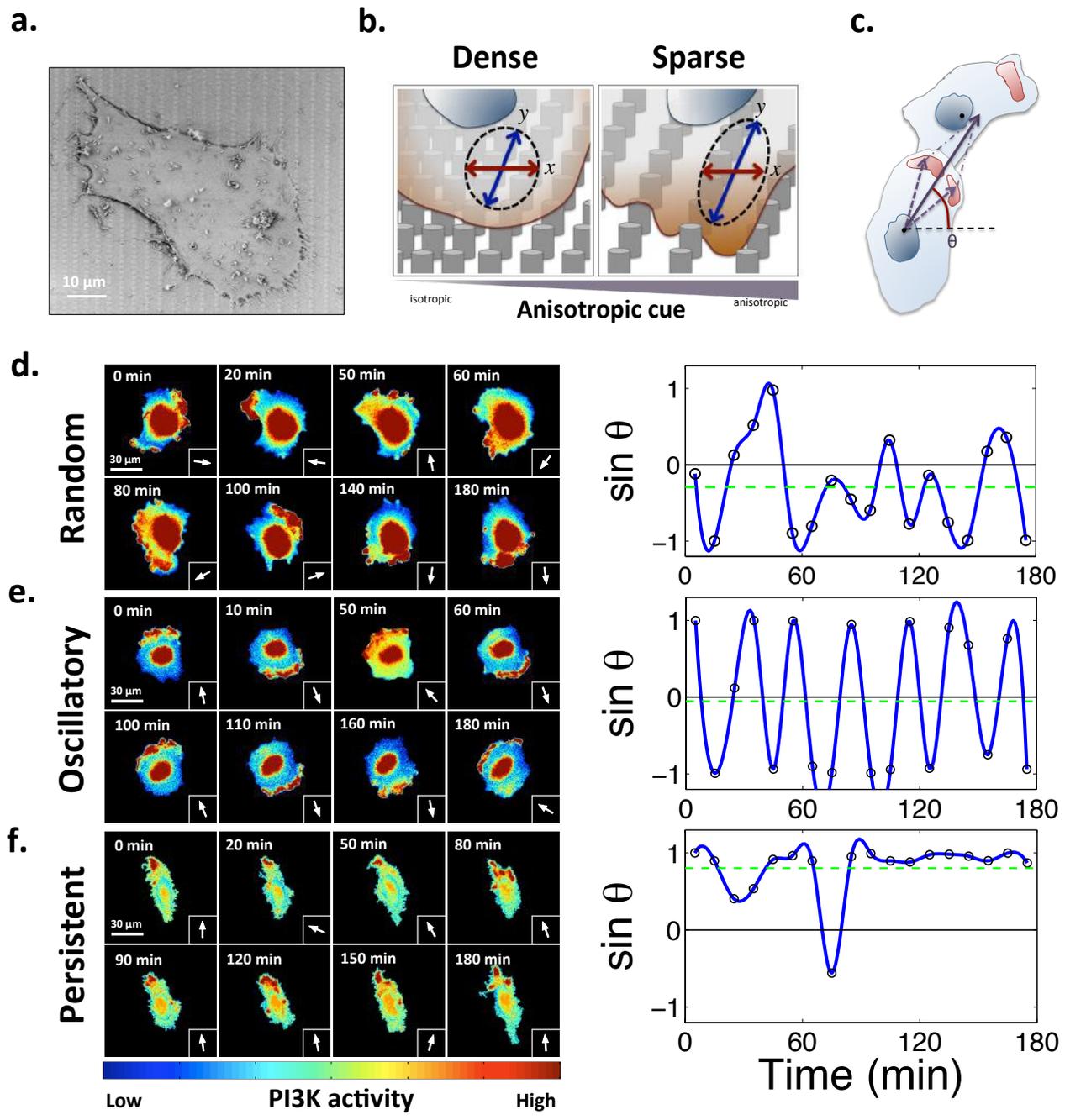

**Figure 1**

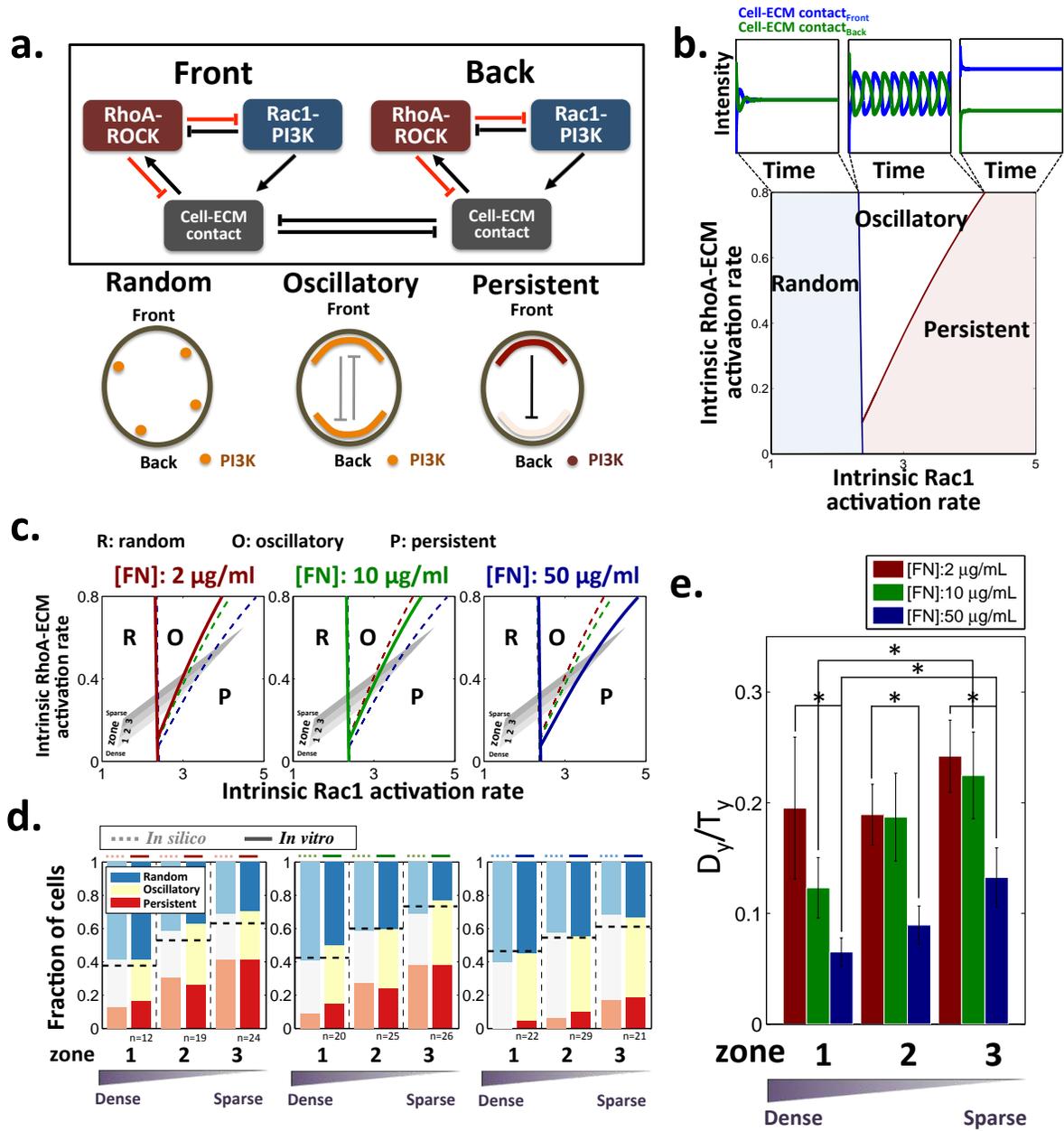

Figure 2

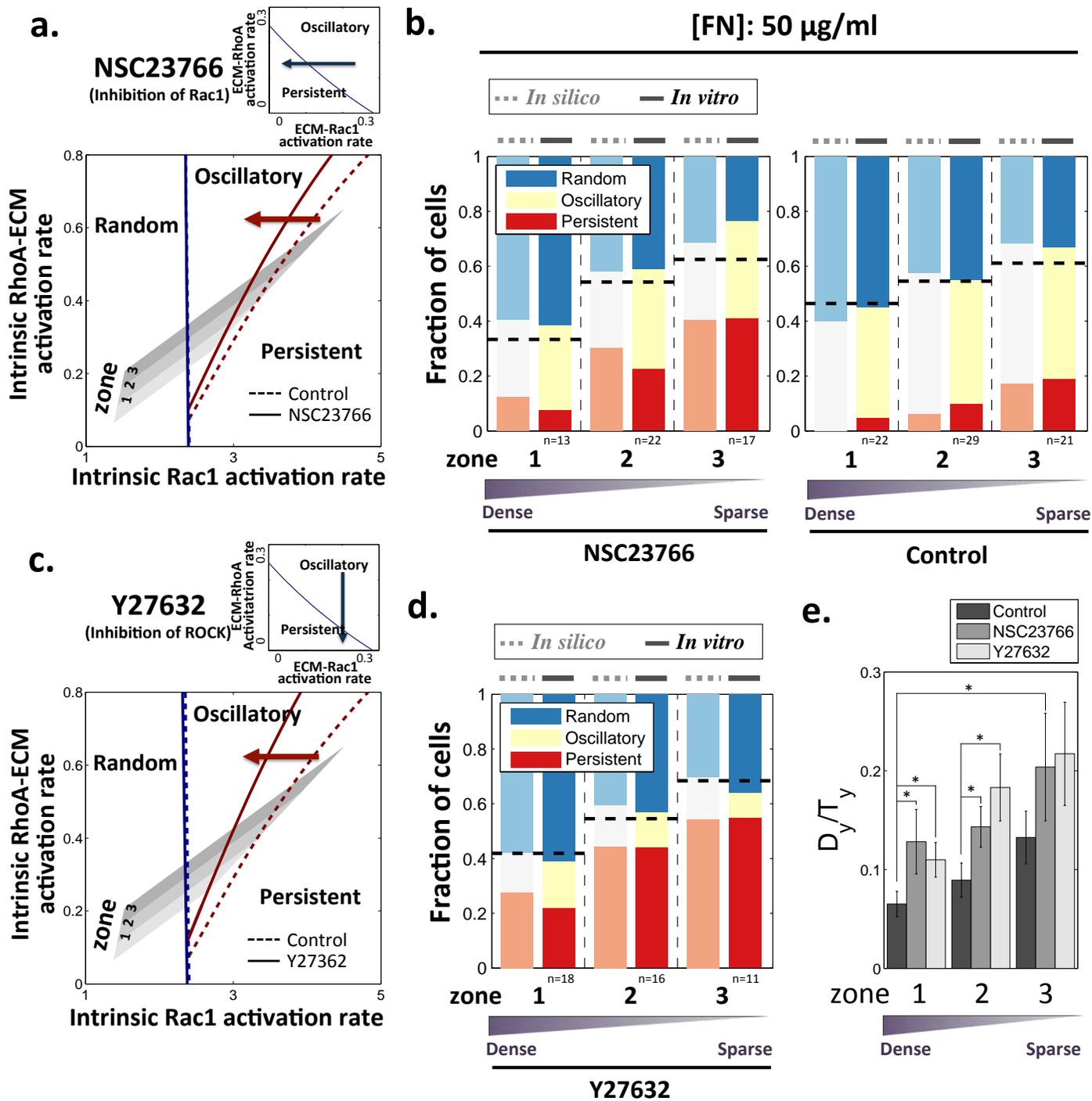

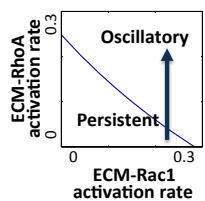

**a.** Nocodazole (Inhibition of microtubule)

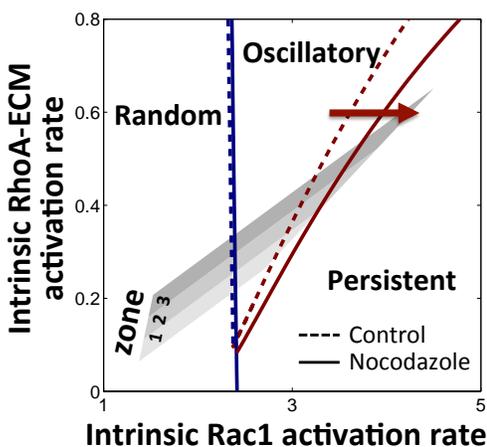

**b.**

Active GEFH1 — 100 kDa
GEFH1 — 100 kDa
GAPDH — 37 kDa

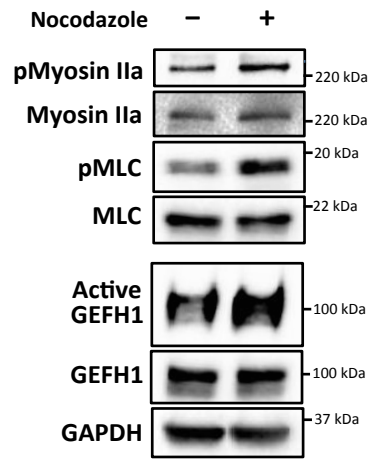

**c.** [FN]: 10 µg/ml

*In silico* ---- **In vitro** ——

- Random
- Oscillatory
- Persistent

Control / Nocodazole
Dense → Sparse

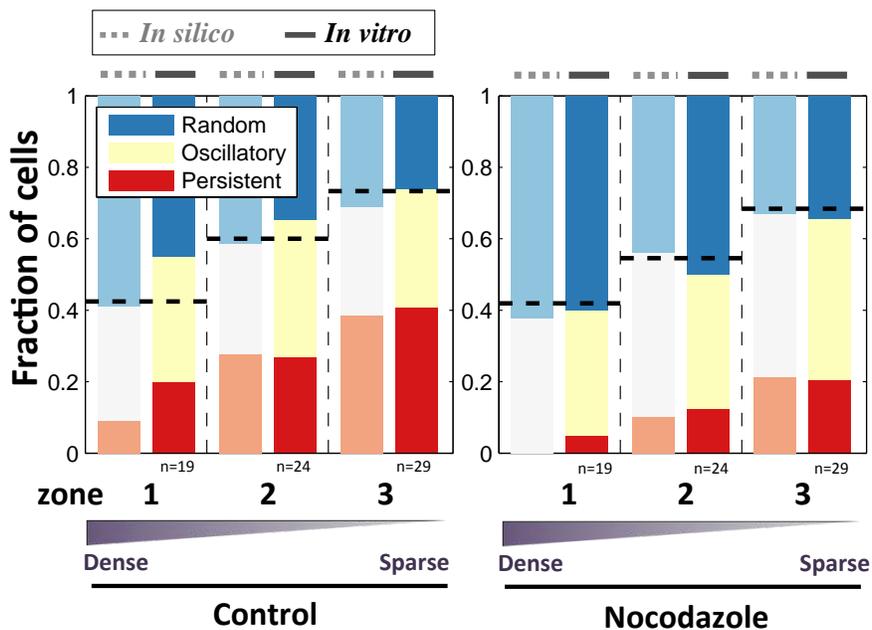
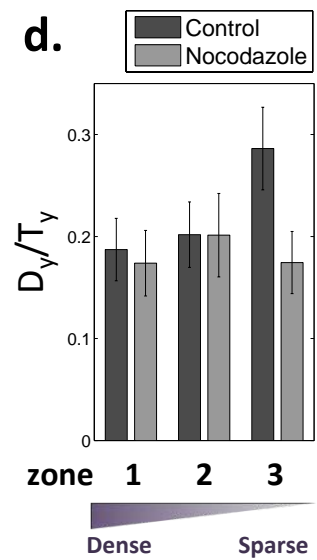

$D_y/T_y$ vs zone

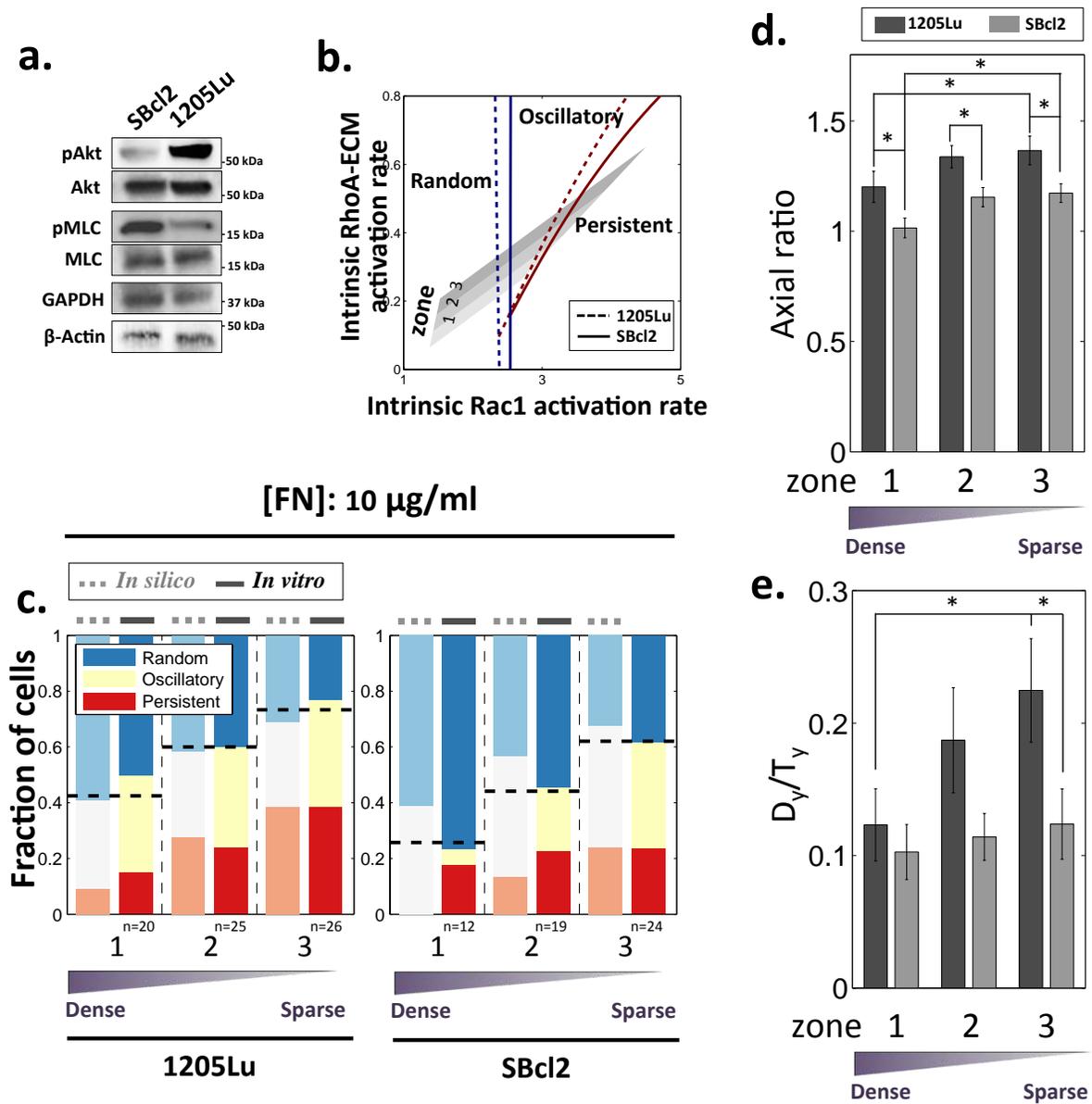

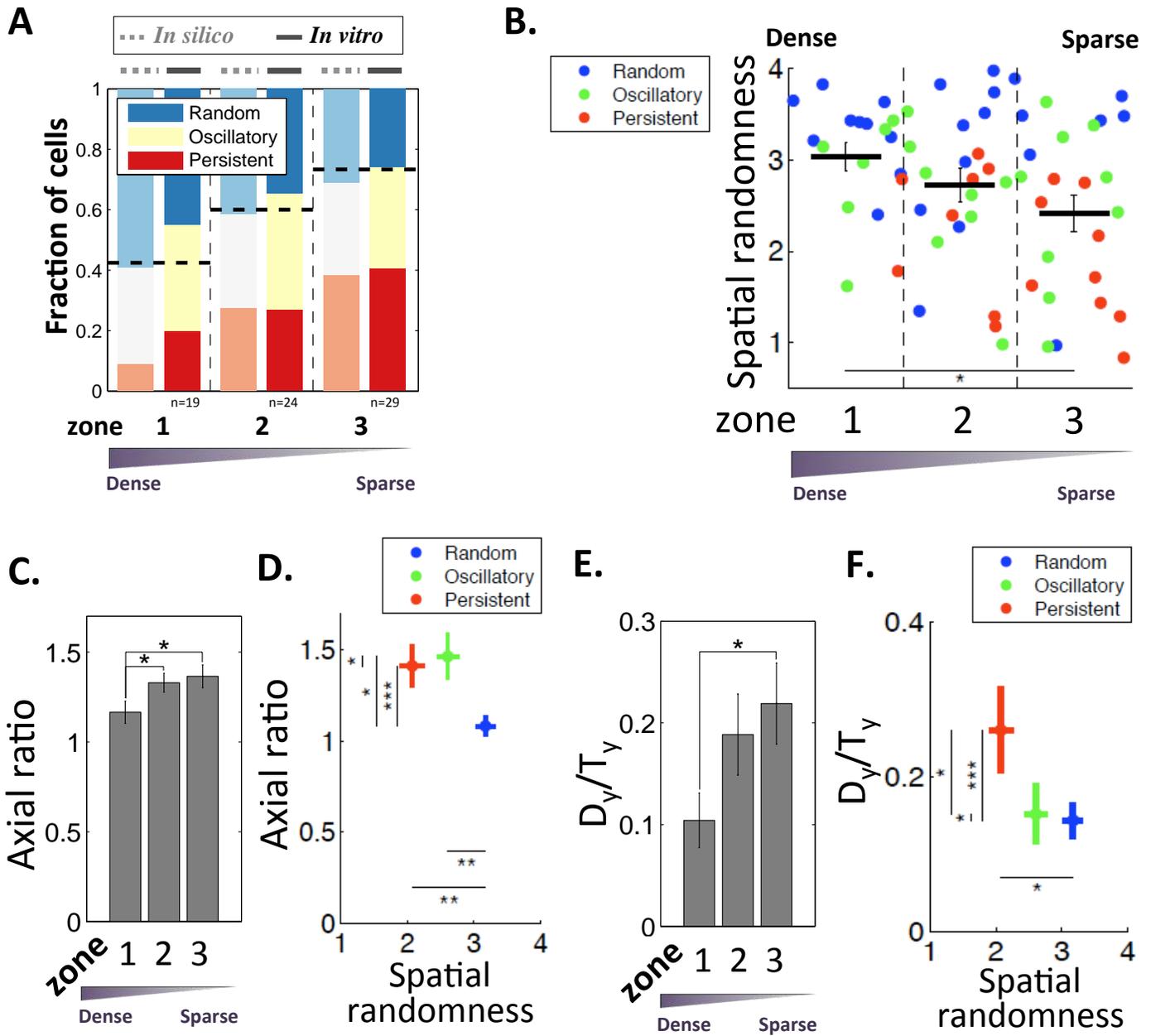

**Supplementary Fig. 1**

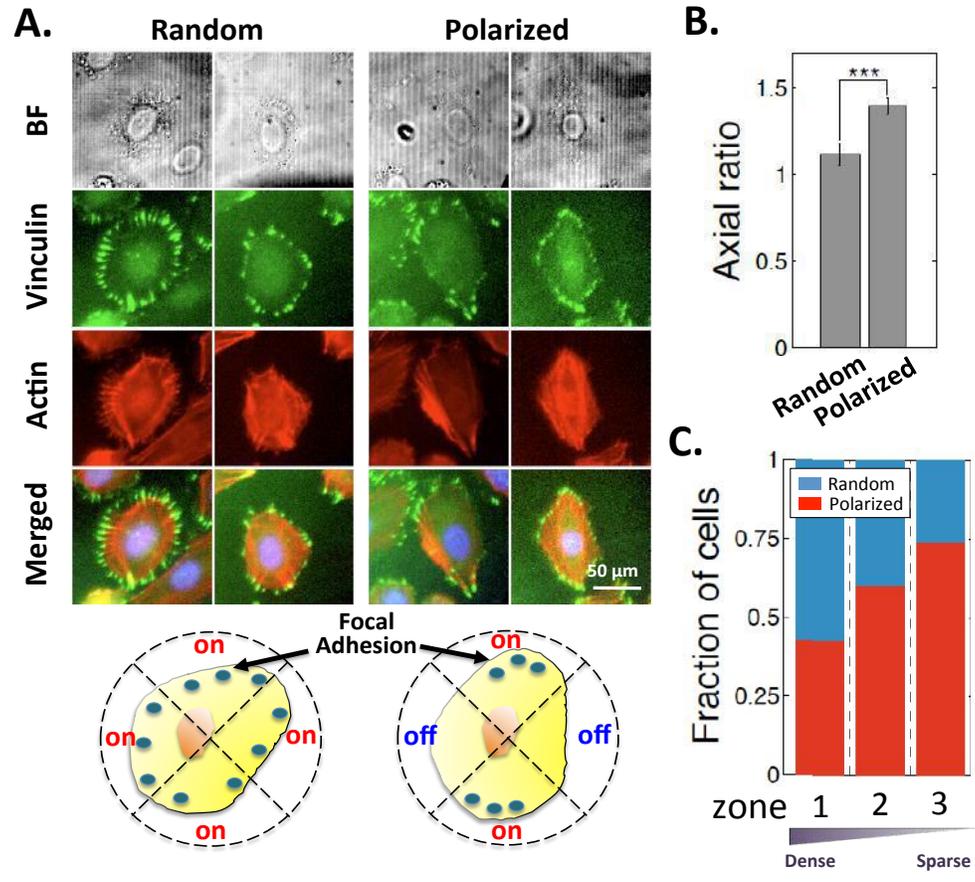

**Supplementary Fig. 2**

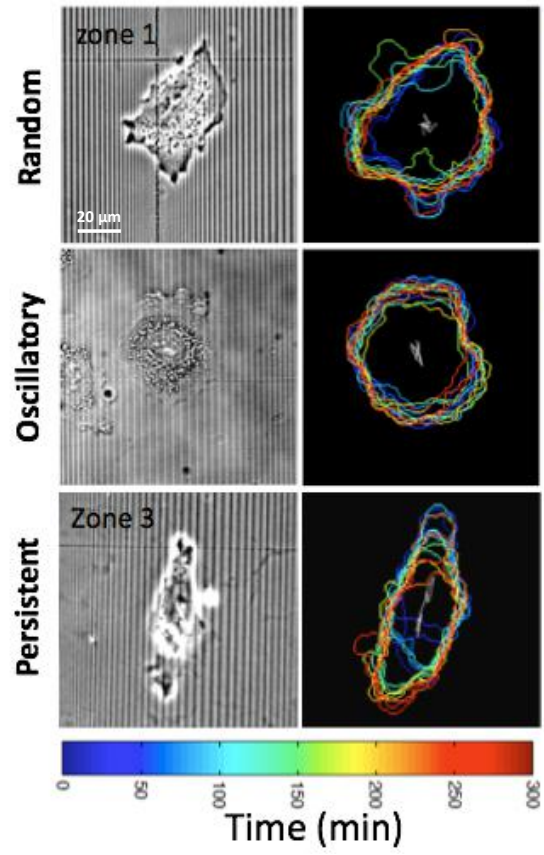

**Supplementary Fig.3**

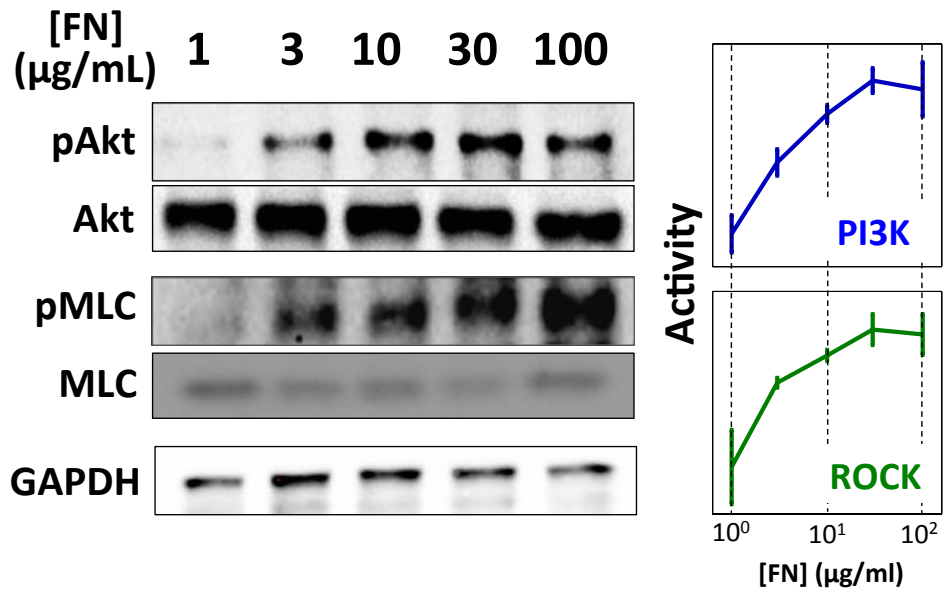

**Supplementary Fig.4**

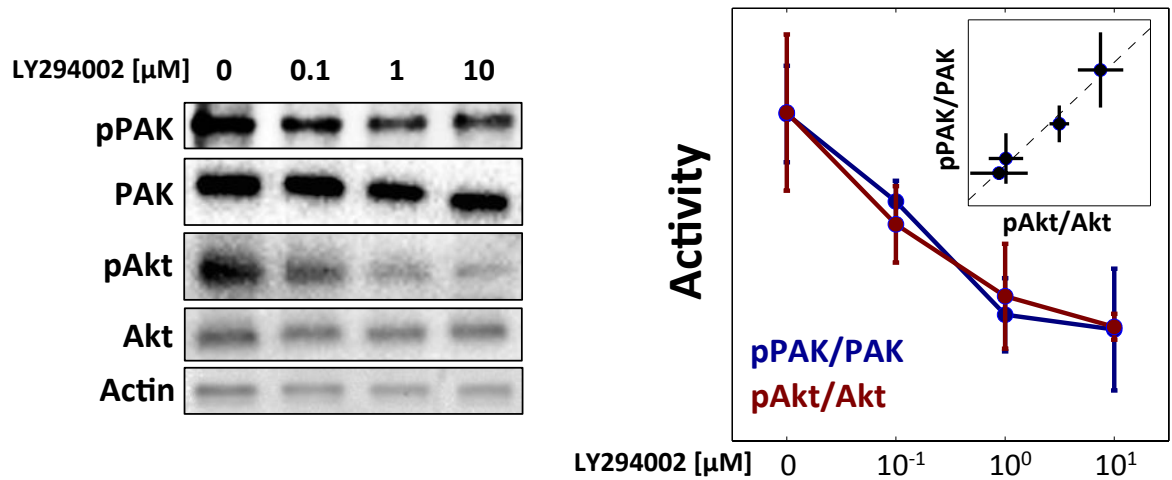

**Supplementary Fig.5**

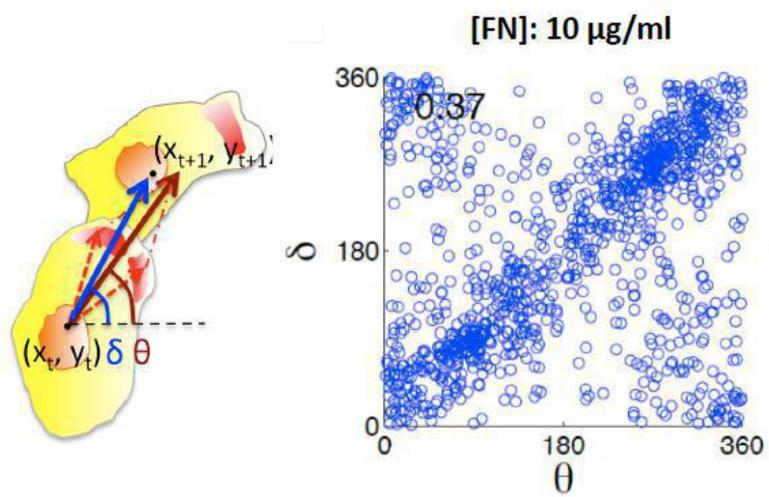

**Supplementary Fig.6**

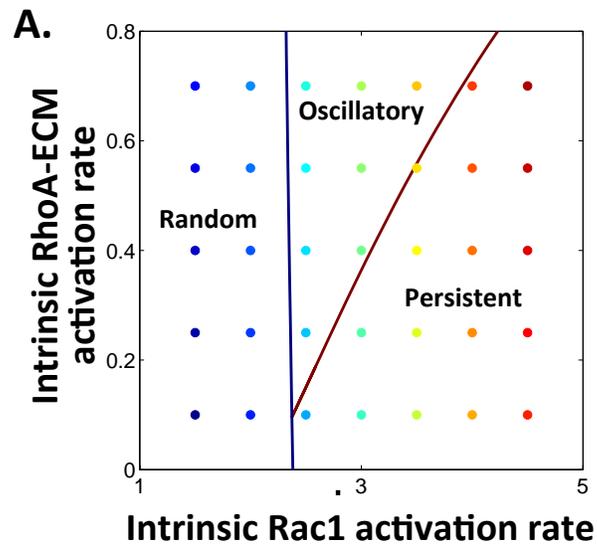

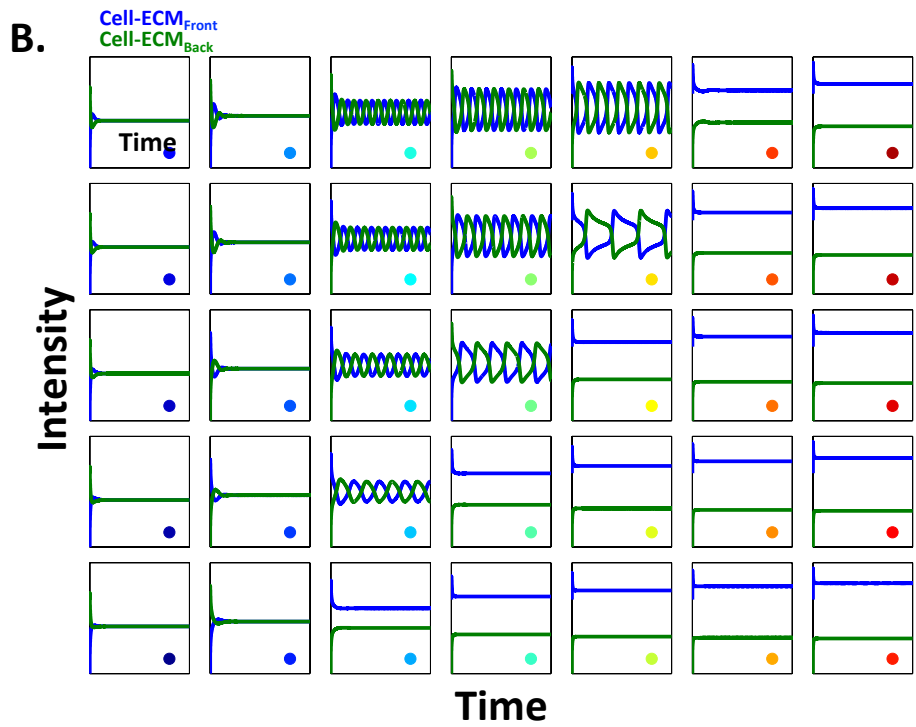

**Supplementary Fig. 7**

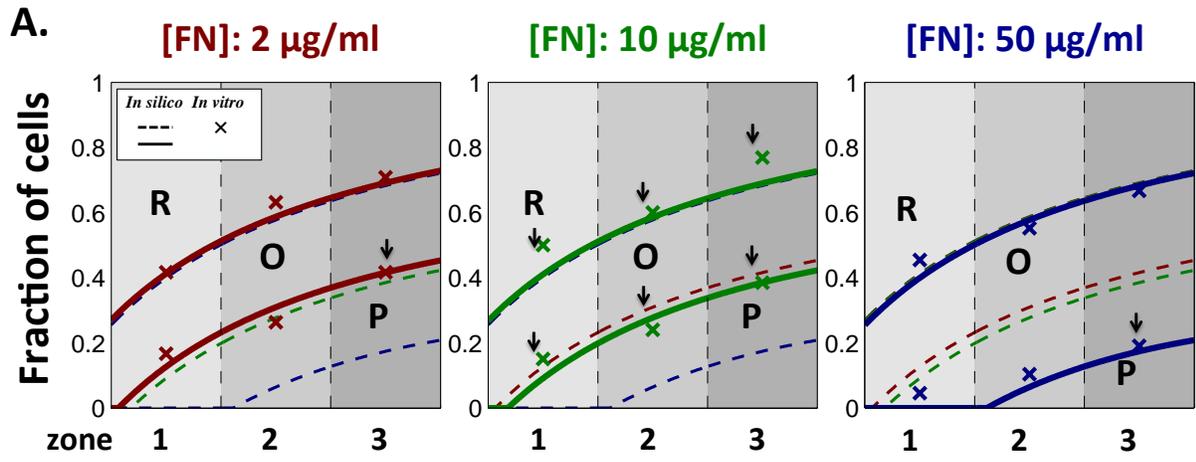
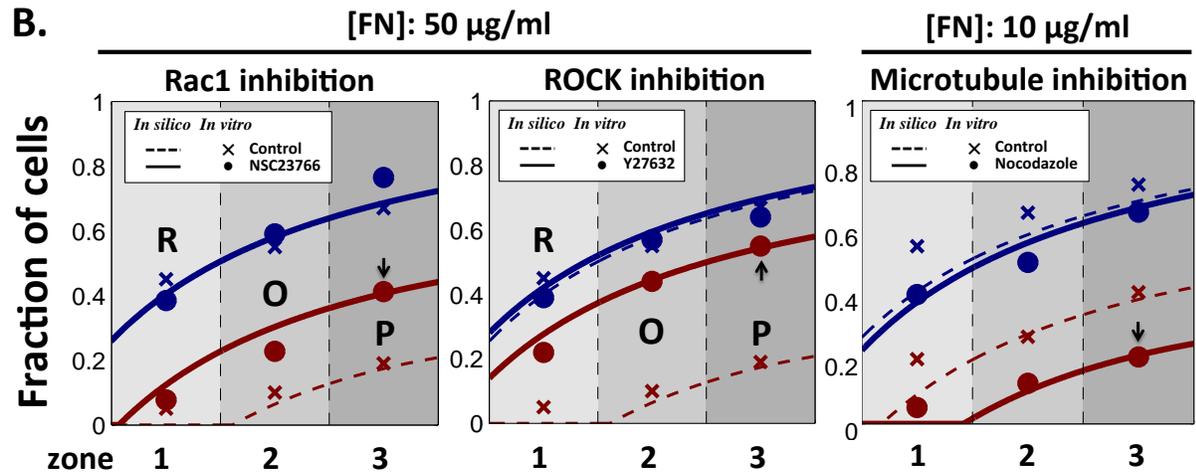
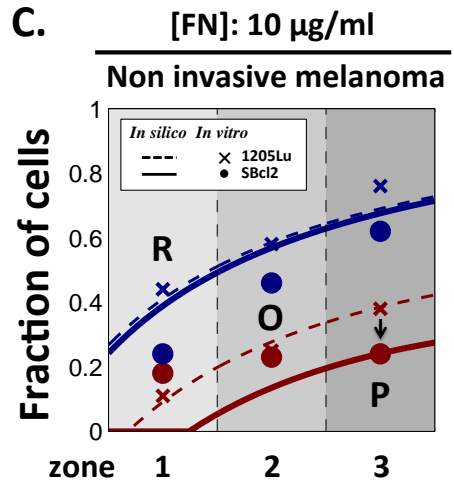

R: random
O: oscillatory
P: persistent

Supplementary Fig.8

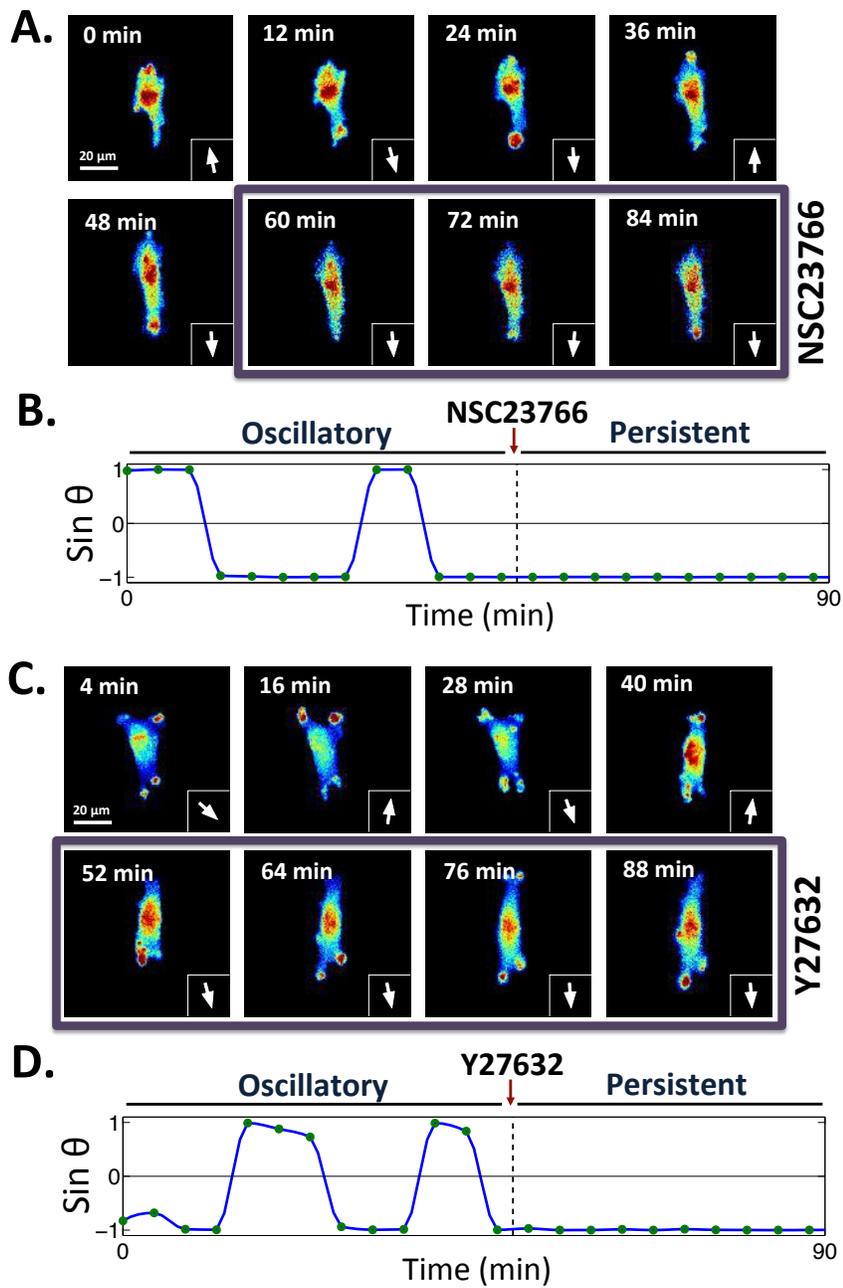

**Supplementary Fig.9**

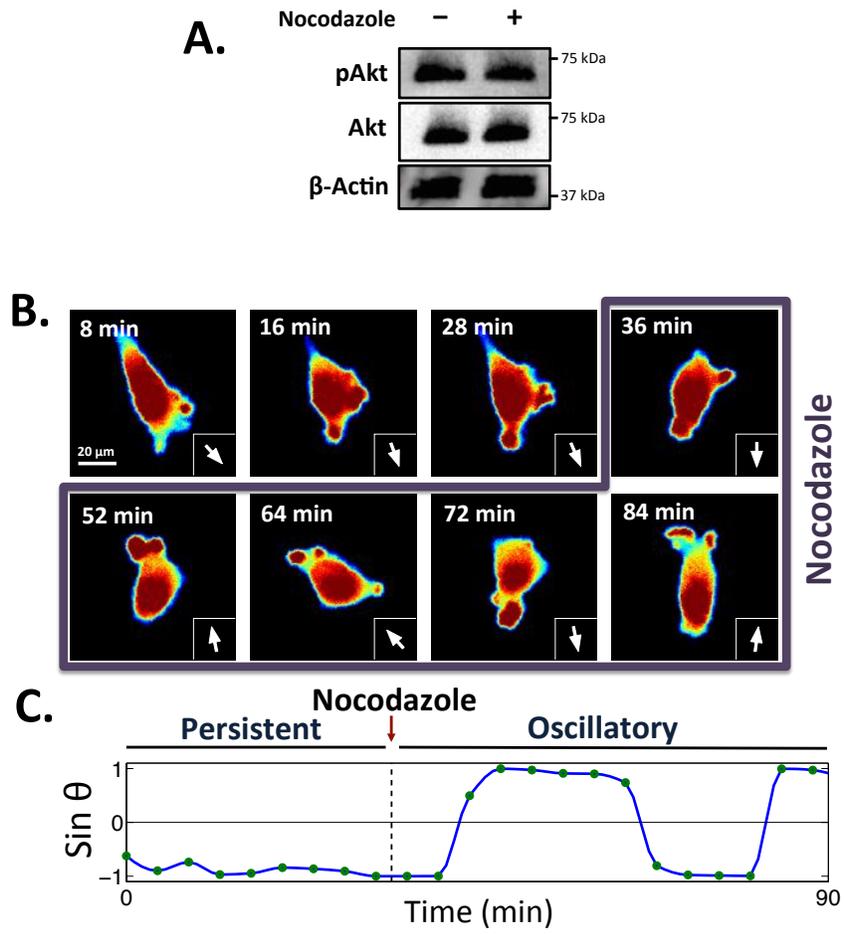

**Supplementary Fig.10**

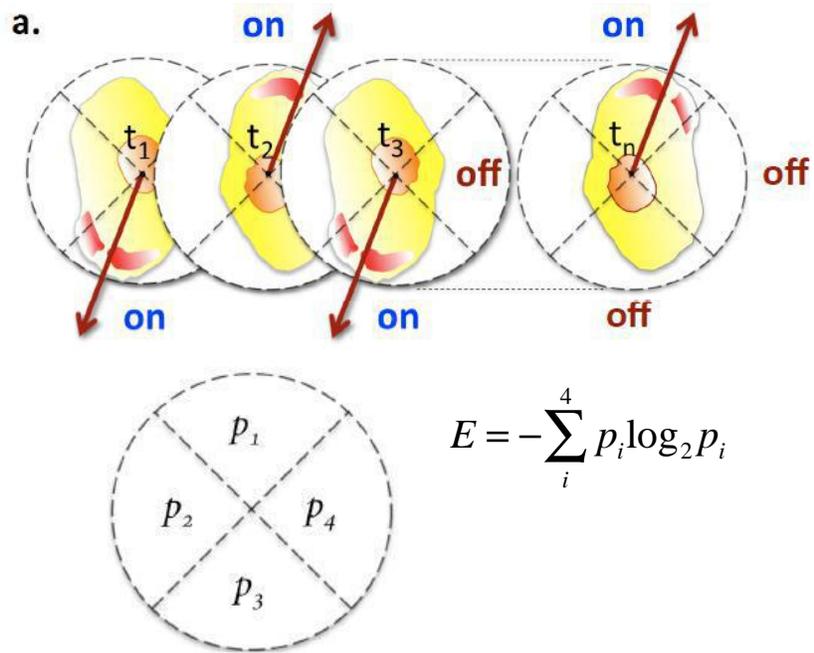

$$E = -\sum_{i}^{4} p_i \log_2 p_i$$

**Supplementary Fig. 11**

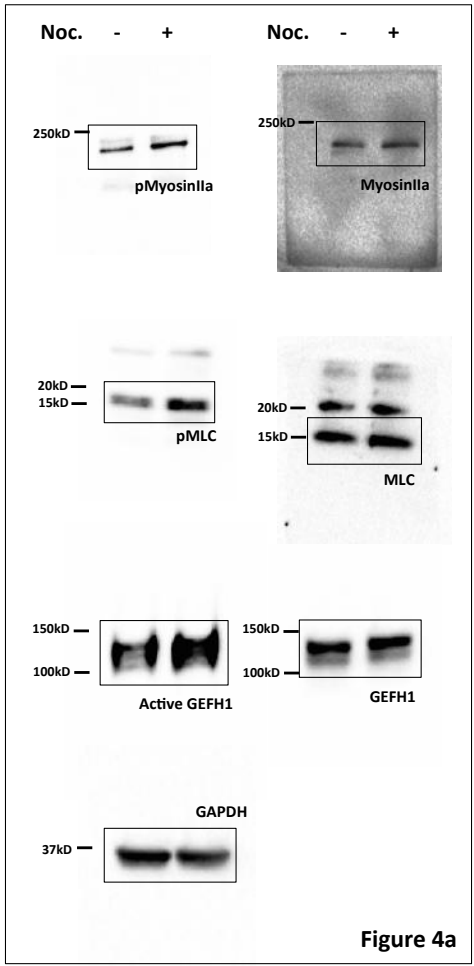

Figure 4a

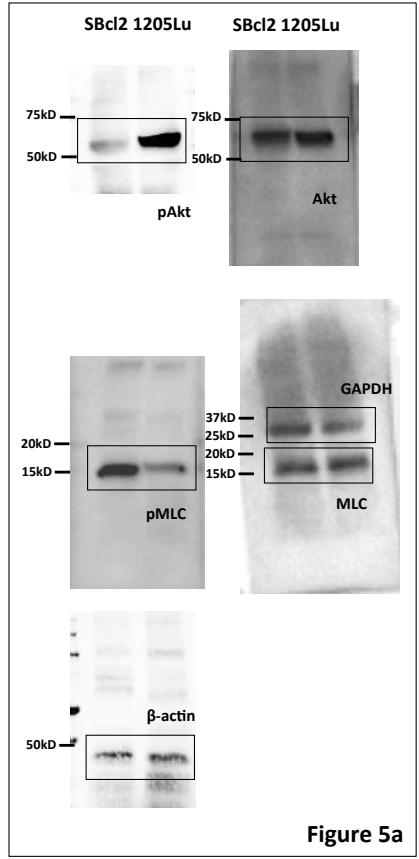

Figure 5a

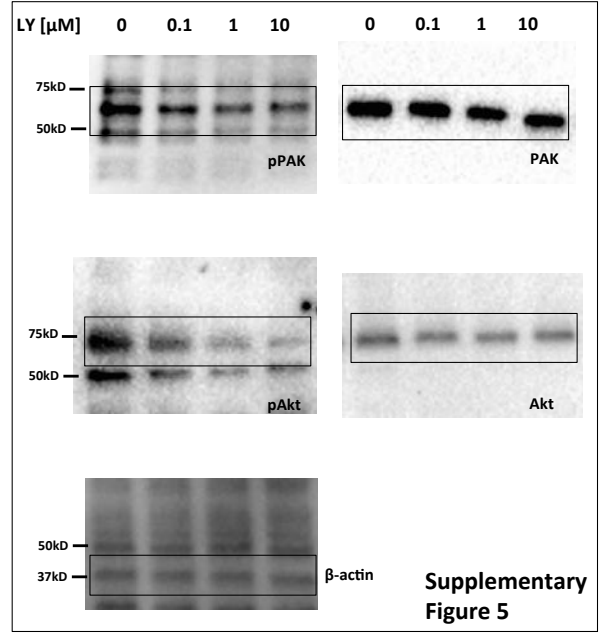

Supplementary Figure 5

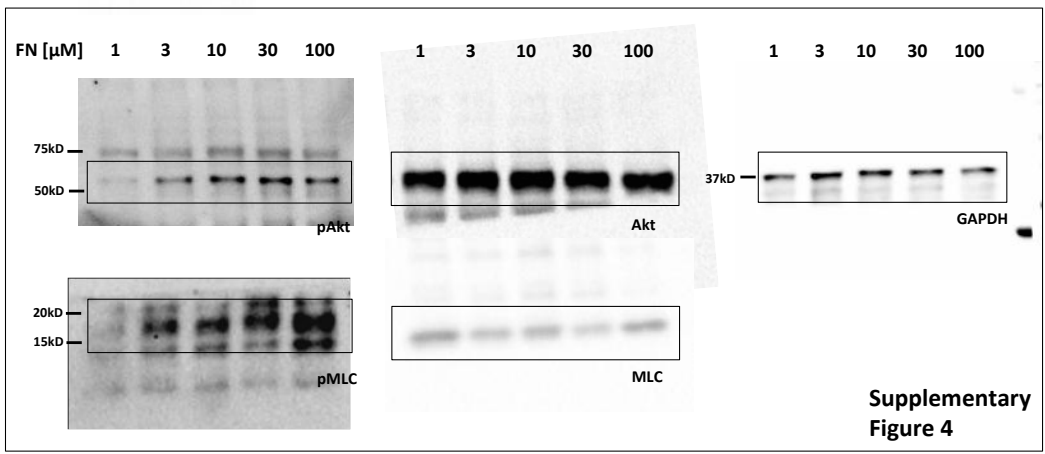

Supplementary Figure 4

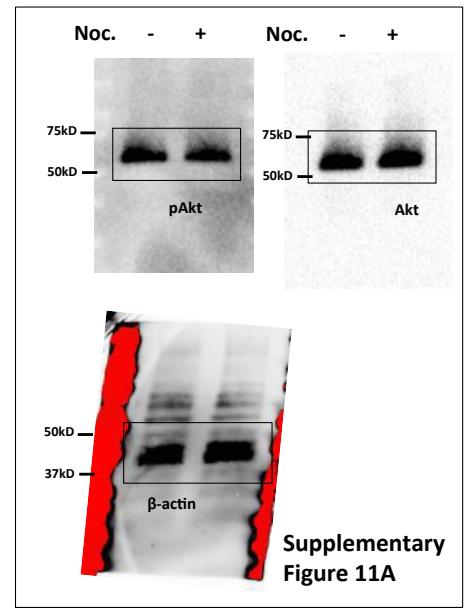

Supplementary Figure 11A

**Supplementary Fig. 12**